\documentclass{article}

\usepackage{report}
\usepackage[utf8]{inputenc}
\usepackage[T1]{fontenc}

\usepackage{amsmath}
\usepackage{amsfonts}
\usepackage{amssymb}
\usepackage{mathtools}
\usepackage{amsthm}
\usepackage{nicefrac}

\usepackage{fvextra}

\usepackage{booktabs}
\usepackage{array}
\usepackage{tabularx}

\usepackage{graphicx}
\usepackage{subcaption}
\usepackage{float}
\usepackage{wrapfig}
\usepackage{adjustbox}
\usepackage{changepage}
\usepackage{tikz}

\usepackage{caption}

\usepackage{array}
\usepackage{booktabs}
\usepackage{longtable}
\usepackage{multirow}
\usepackage{makecell}
\usepackage{tabularx}
\usepackage{colortbl}
\usepackage{arydshln}

\usepackage{algorithm}
\usepackage{algpseudocode}
\usepackage{listings}

\usepackage{enumitem}
\usepackage{microtype}
\usepackage{lipsum}
\usepackage{xspace}
\usepackage{pifont}
\usepackage{fontawesome5}
\usepackage{titlesec}
\usepackage{tocloft}

\usepackage{xcolor}
\usepackage{etoolbox}
\usepackage[most]{tcolorbox}
\usepackage{bibunits}

\usepackage{hyperref}
\usepackage{cleveref}

\usepackage{graphicx}
\usepackage{booktabs}
\usepackage{tabularx}
\usepackage{array}

\newcommand{\modellogo}[1]{%
  \raisebox{-0.18\height}{\includegraphics[height=1em]{Fig/model_logo/#1}}%
}

\definecolor{odMuted}{HTML}{737D8C}
\definecolor{odOverall}{HTML}{F0F4F8}
\definecolor{odAccent}{HTML}{A64046}

\newcommand{\odacc}[2]{%
  #1\,{\scriptsize\textcolor{odMuted}{(#2)}}%
}
\newcommand{\odbest}[2]{%
  \textbf{#1}\,{\scriptsize\textcolor{odMuted}{(#2)}}%
}
\newcommand{\odtasks}[1]{%
  {\scriptsize\textcolor{odMuted}{#1 tasks}}%
}

\titlespacing{\paragraph}{0pt}{0.35em}{0.6em}

\usepackage{amsmath,amsfonts,bm}

\def\eqref#1{equation~\ref{#1}}

\def\1{\bm{1}}

\DeclareMathAlphabet{\mathsfit}{\encodingdefault}{\sfdefault}{m}{sl}
\SetMathAlphabet{\mathsfit}{bold}{\encodingdefault}{\sfdefault}{bx}{n}

\definecolor{cornellred}{rgb}{0.7, 0.11, 0.11}
\definecolor{cadmiumgreen}{rgb}{0.0, 0.42, 0.24}
\definecolor{aliceblue}{rgb}{0.91, 0.94, 0.97}
\definecolor{darkblue}{rgb}{0.83, 0.89, 0.97}

\definecolor{Red7}{rgb}{0.941, 0.243, 0.243}
\definecolor{Green7}{RGB}{55, 178, 77}
\definecolor{Blue9}{rgb}{0.098, 0.3, 0.9}
\definecolor{SJViolet}{RGB}{105, 100, 171}
\definecolor{SJRed}{RGB}{237, 109, 107}

\definecolor{DojoBlue}{HTML}{498AE5}
\definecolor{LinkBlue}{HTML}{0000FF}
\definecolor{Blue_1}{HTML}{1C6EF3}
\definecolor{Blue_2}{HTML}{D2E3FC}
\definecolor{Dojo_Yellow}{HTML}{FFDF7B}

\definecolor{AbsBlue}{HTML}{498AE5}
\definecolor{AbsCyan}{HTML}{23B399}

\newsavebox{\roundedimgbox}

\usepackage[most]{tcolorbox}
\usepackage{listings}

\newtcblisting{promptbox}[1]{
    enhanced,
    breakable,
    listing only,
    listing engine=listings,
    colback=black!1,
    colframe=black!22,
    colbacktitle=black!5,
    coltitle=black!85,
    fonttitle=\bfseries\footnotesize,
    title={#1},
    boxrule=0.4pt,
    arc=1.5pt,
    left=9pt,
    right=9pt,
    top=8pt,
    bottom=8pt,
    toptitle=5pt,
    bottomtitle=5pt,
    borderline west={2pt}{0pt}{black!70},
    before skip=8pt,
    after skip=10pt,
    listing options={
        basicstyle=\ttfamily\scriptsize,
        breaklines=true,
        breakatwhitespace=true,
        columns=fullflexible,
        keepspaces=true,
        showstringspaces=false,
        upquote=true
    }
}

\newcolumntype{C}[1]{%
  >{\centering\arraybackslash}m{#1}%
}

\newcolumntype{L}[1]{%
  >{\raggedright\arraybackslash}m{#1}%
}

\newsavebox{\taskimgbox}

\newif\ifTaskDataSourcePending
\newcommand{\TaskDataSourceText}{}

\newcommand{\flushtaskdatasource}{%
  \ifTaskDataSourcePending
    \noindent
    \TaskDataSourceText
    \par
    \vspace{0.12em}%
    \TaskDataSourcePendingfalse
  \fi
}

\newcommand{\taskitem}[2]{%
  \ifstrequal{#1}{Data Source}{%
    \gdef\TaskDataSourceText{\textbf{#1}: #2}%
    \TaskDataSourcePendingtrue
  }{%
    \ifstrequal{#1}{Usage}{%
      \ifTaskDataSourcePending
        \noindent
        \TaskDataSourceText
        \hspace{2em}
        \textbf{#1}: #2
        \par
        \vspace{0.12em}%
        \TaskDataSourcePendingfalse
      \else
        \noindent
        \textbf{#1}: #2
        \par
        \vspace{0.12em}%
      \fi
    }{%
      \flushtaskdatasource
      \noindent
      \textbf{#1}: #2
      \par
      \vspace{0.12em}%
    }%
  }%
}

\definecolor{agentrow}{HTML}{F0F6EE}
\definecolor{modelrow}{HTML}{FFF6E8}
\definecolor{harnessrow}{HTML}{EEF5FB}
\definecolor{termrow}{HTML}{F5F0FA}

\definecolor{lastexamred}{HTML}{D00000}

\definecolor{sc0}{HTML}{9C2F2F}
\definecolor{sc1}{HTML}{B45F06}
\definecolor{sc2}{HTML}{8A6D1D}
\definecolor{sc3}{HTML}{4F7C2C}
\definecolor{sc4}{HTML}{2E6F3E}
\definecolor{sc5}{HTML}{1D5B35}

\newcommand{\hsc}[1]{%
  \ifdim #1 pt < 5pt
    \textcolor{sc0}{#1}%
  \else\ifdim #1 pt < 20pt
    \textcolor{sc1}{#1}%
  \else\ifdim #1 pt < 40pt
    \textcolor{sc2}{#1}%
  \else\ifdim #1 pt < 60pt
    \textcolor{sc3}{#1}%
  \else\ifdim #1 pt < 80pt
    \textcolor{sc4}{#1}%
  \else
    \textcolor{sc5}{#1}%
  \fi\fi\fi\fi\fi
}

\newcommand{\hscb}[1]{%
  \ifdim #1 pt < 5pt
    \textcolor{sc0}{\textbf{#1}}%
  \else\ifdim #1 pt < 20pt
    \textcolor{sc1}{\textbf{#1}}%
  \else\ifdim #1 pt < 40pt
    \textcolor{sc2}{\textbf{#1}}%
  \else\ifdim #1 pt < 60pt
    \textcolor{sc3}{\textbf{#1}}%
  \else\ifdim #1 pt < 80pt
    \textcolor{sc4}{\textbf{#1}}%
  \else
    \textcolor{sc5}{\textbf{#1}}%
  \fi\fi\fi\fi\fi
}

\lstdefinestyle{compactpython}{
  language=Python,
  basicstyle=\ttfamily\scriptsize,
  keywordstyle=\color{blue!70!black}\bfseries,
  commentstyle=\color{green!40!black},
  stringstyle=\color{orange!80!black},
  numbers=none,
  frame=single,
  framerule=0.4pt,
  rulecolor=\color{black!20},
  backgroundcolor=\color{gray!3},
  breaklines=true,
  breakatwhitespace=true,
  showstringspaces=false,
  tabsize=2,
  columns=fullflexible,
  keepspaces=true,
  aboveskip=3pt,
  belowskip=3pt,
  xleftmargin=0pt,
  framexleftmargin=3pt,
  framexrightmargin=3pt,
  framesep=3pt,
  lineskip=-1pt
}

\hypersetup{
  colorlinks=true,
  linkcolor=LinkBlue,
  citecolor=LinkBlue,
  urlcolor=LinkBlue
}

\colorlet{rlTitle}{rlwrldBlack}
\colorlet{rlSectionNum}{Blue_1}
\colorlet{rlHyperlink}{rlwrldDarkGreen}
\colorlet{rlRefLink}{rlwrldDarkGreen}
\colorlet{rlTocLink}{rlwrldBlack}

\def\abstractstyle{background}
\def\abstractposition{intro}

\def\authordisplay{full}

\def\logotype{woMark}

\title{\centering Deep Search in Real-World Web Environments}

\ifdefstring{\authordisplay}{full}{%

  \author{%

    \normalsize
    \linespread{1.25}\selectfont
    \centering

    {\small

      \mbox{%
        \textbf{Xiaomin Yu}\textsuperscript{1\S\ddag}%
        \hspace{0.5em}%
        \textbf{Hongliang Chen}\textsuperscript{1\ddag}%
        \hspace{0.5em}%
        Zhe Huang\textsuperscript{1\ddag}%
        \hspace{0.5em}%
        Shilin Yan\textsuperscript{2\S\ddag}%
        \hspace{0.5em}%
        Shao-Rong Su\textsuperscript{3\ddag}%
        \hspace{0.5em}%
        Xuqiao Ma\textsuperscript{1\ddag}%
        \hspace{0.5em}%
        Feiyang Wang\textsuperscript{4\ddag}%
        \hspace{0.5em}%
        Jiayu Chen\textsuperscript{1\ddag}%
      }%
      \par

      \mbox{%
        Jiwei Fang\textsuperscript{1\ddag}%
        \hspace{0.5em}%
        Yu Yan\textsuperscript{1\ddag}%
        \hspace{0.5em}%
        Yini Huang\textsuperscript{1\ddag}%
        \hspace{0.5em}%
        Chen Chen\textsuperscript{5\ddag}%
        \hspace{0.5em}%
        Ruoxi Cheng\textsuperscript{2\ddag}%
        \hspace{0.5em}%
        Yuzhi He\textsuperscript{1\ddag}%
        \hspace{0.5em}%
        Yuhui Zhang\textsuperscript{6}%
        \hspace{0.5em}%
        Xiaoxing Hu\textsuperscript{7\ddag}%
        \hspace{0.5em}%
        Chen Liu\textsuperscript{8}%
      }%
      \par

      \mbox{%
        Jianbo Lin\textsuperscript{9}%
        \hspace{0.5em}%
        Yijiang Li\textsuperscript{10}%
        \hspace{0.5em}%
        Xiaobin Hu\textsuperscript{11}%
        \hspace{0.5em}%
        Menglong Li\textsuperscript{1}%
        \hspace{0.5em}%
        Zhuoran Lu\textsuperscript{12}%
        \hspace{0.5em}%
        Weishi Wang\textsuperscript{13}%
        \hspace{0.5em}%
        Zhongqi Yue\textsuperscript{14}%
        \hspace{0.5em}%
        Chengwei Qin\textsuperscript{1\S\dag}%
      }%
      \par
    }

    {\footnotesize
      \resizebox{\textwidth}{!}{%
        \mbox{%
          \textsuperscript{1} HKUST(GZ)
          \quad
          \textsuperscript{2} FDU
          \quad
          \textsuperscript{3} UW
          \quad
          \textsuperscript{4} IC
          \quad
          \textsuperscript{5} NTU
          \quad
          \textsuperscript{6} Stanford
          \quad
          \textsuperscript{7} SJTU
          \quad
          \textsuperscript{8} Yale
          \quad
          \textsuperscript{9} NJU
          \quad
          \textsuperscript{10} UCSD
          \quad
          \textsuperscript{11} NUS
          \quad
          \textsuperscript{12} HKU
          \quad
          \textsuperscript{13} SAP
          \quad
          \textsuperscript{14} Microsoft Research
        }%
      }%
      \par
    }

    {\footnotesize

      \textsuperscript{\S}Project Leaders
      \quad
      \textsuperscript{\ddag}Main Contributors
      \quad
      \textsuperscript{\dag}Corresponding Authors

      \par
    }

  }%

}{}

\AddToHook{shipout/foreground}{%
  \ifnum\value{page}=1
    \begin{tikzpicture}[remember picture,overlay]
      \node[
        anchor=north west,
        inner sep=0pt,
        xshift=1.8cm,
        yshift=-1.2cm
      ] at (current page.north west) {%
        \includegraphics[
          width=5.0cm,
          height=0.8cm,
          keepaspectratio
        ]{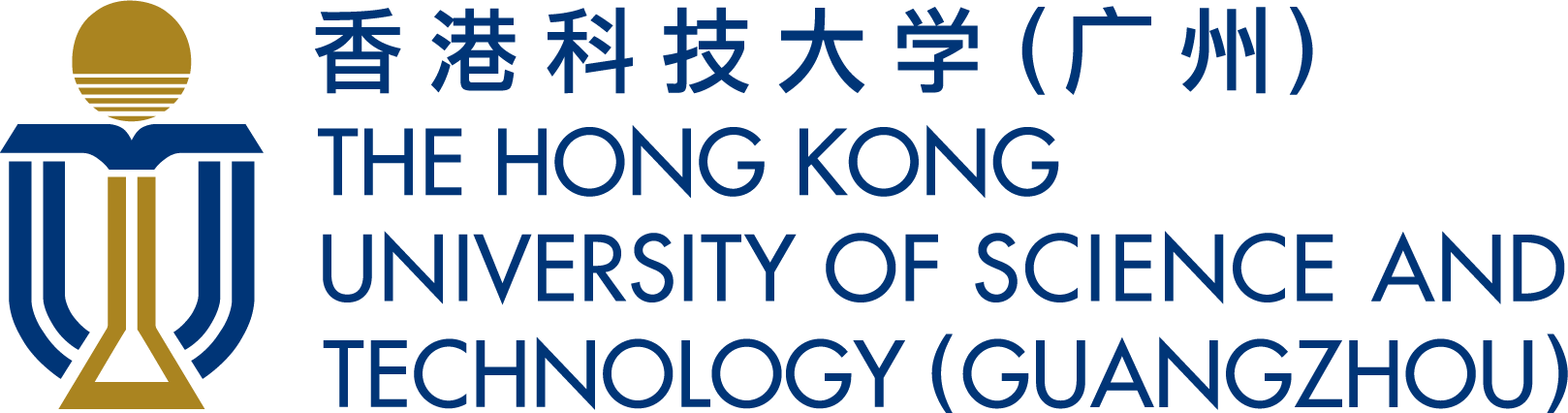}%
      };
    \end{tikzpicture}%
  \fi
}

\begin{document}

\maketitle

\begingroup

\noindent
\makebox[\textwidth][c]{%
\begin{tabular}{
  @{}
  c
  @{\hspace{1.1em}}
  c
  @{\hspace{1.1em}}
  c
  @{\hspace{1.1em}}
  c
  @{}
}

\raisebox{-1.5pt}{%
  \includegraphics[height=1.05em]{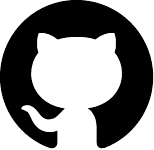}%
}%
~\href{https://github.com/Yu-xm/OpenDS}{%
  \texttt{Code}%
}

&

\raisebox{-1.5pt}{%
  \includegraphics[height=1.05em]{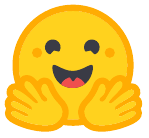}%
}%
~\href{https://huggingface.co/datasets/Yu2020/OpenDS-World}{%
  \textnormal{Dataset}%
}

&

\raisebox{-1.5pt}{%
  \includegraphics[height=1.05em]{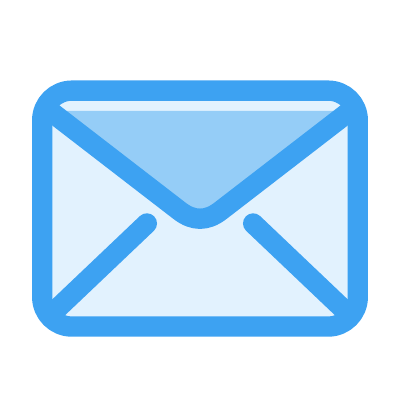}%
}%
~\href{mailto:yuxm02@gmail.com}{%
  \textnormal{Contact}%
}

\end{tabular}%
}

\endgroup

\begin{figure}[h]
    \centering
    \includegraphics[width=0.9\textwidth]{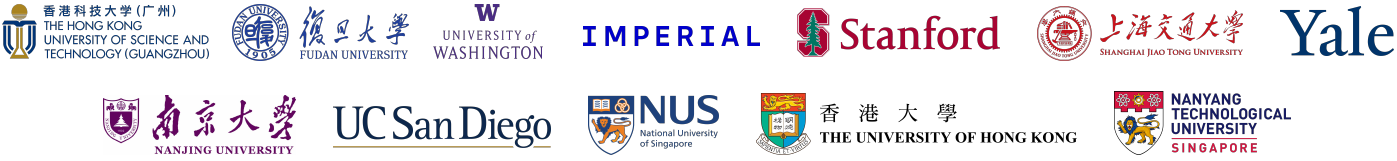}
\end{figure}

\vspace{-3pt}

\begin{center}
    \includegraphics[width=0.89\textwidth]{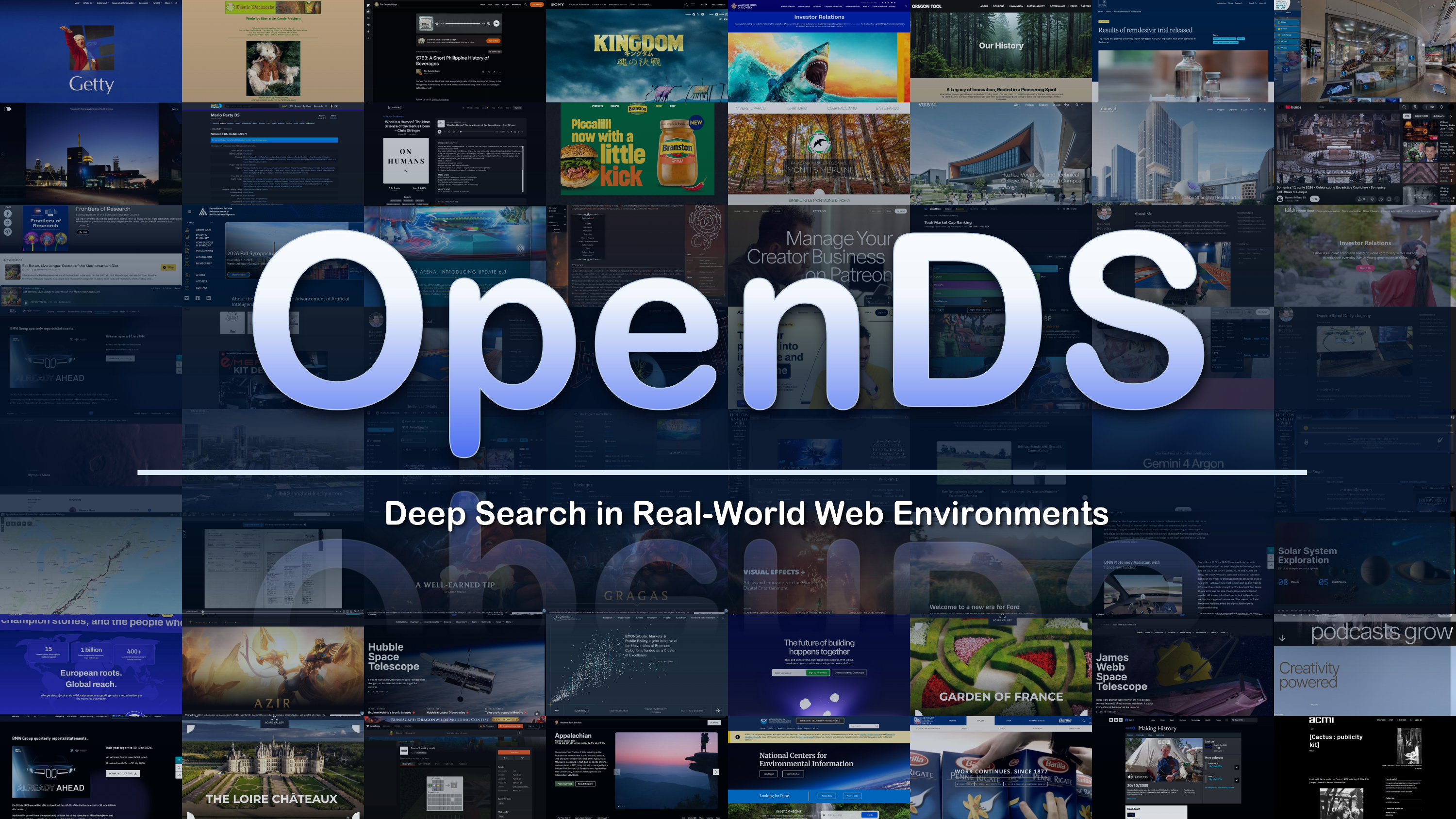}
\end{center}

\vspace{-3pt}

\begin{center}
    \includegraphics[width=0.89\textwidth]{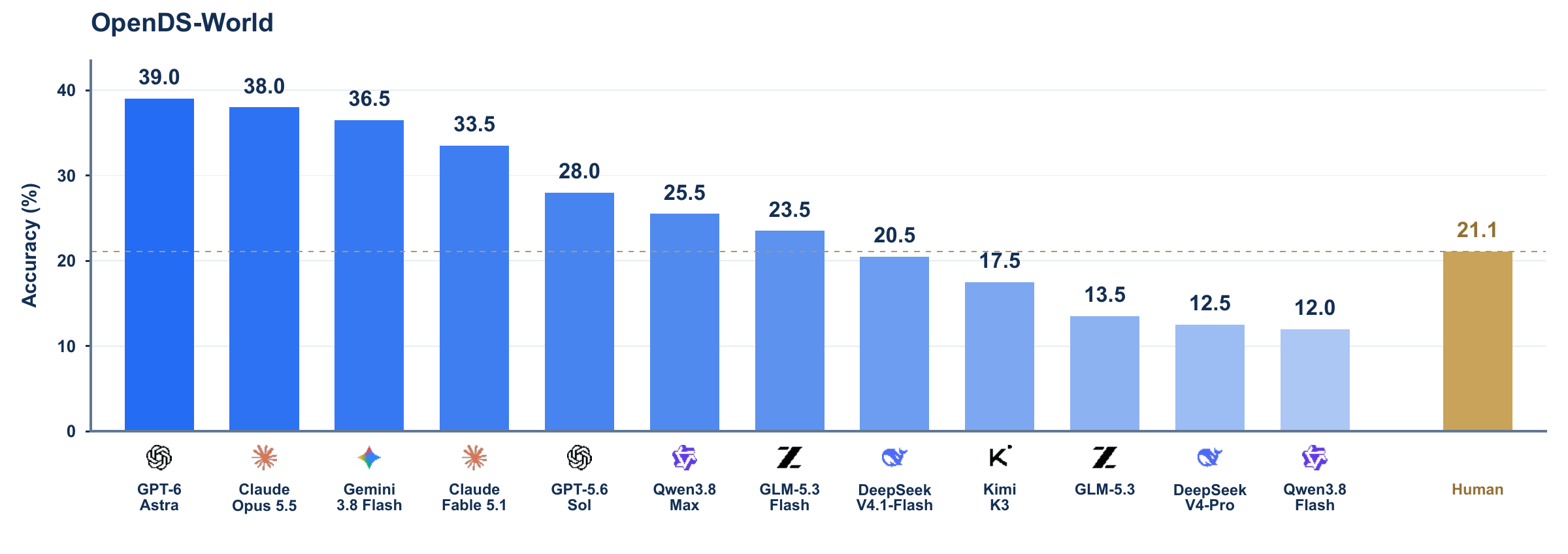}
    \captionof{figure}{\textbf{Model and human performance on OpenDS-World.}}
    \label{fig:opends}
\end{center}


\makeatletter

\newcommand{\rlfootertext}[1]{%
  \g@addto@macro\rl@pagefootnotes{%
    \par
    \noindent
    #1%
  }%
}

\makeatother

\rlfootertext{}


\ifdefstring{\abstractposition}{page1}{%
  \begin{abstract}

Despite substantial progress in Deep Search agents, their ability to conduct end-to-end research in real-world open Web environments remains insufficiently evaluated. Unlike static retrieval settings, information on the real-world Web is distributed across diverse sources, modalities, and heterogeneous formats, often involving complex interaction mechanisms that require agents to autonomously search for, access, and integrate relevant evidence during execution. To address this gap, we introduce the \textbf{OpenDS} task, designed to evaluate agents' ability to perform multi-step deep search in the live, open Web. This task requires agents to autonomously explore the real Web, discover and connect relevant evidence from multiple independent sources through Web search, browser interaction, and evidence processing, and ultimately determine a unique factual answer. Building on this task formulation, we construct \textbf{OpenDS-World}, a benchmark comprising 200 challenging tasks spanning diverse information modalities, heterogeneous file formats, and interactive access requirements, with independent human verification to ensure evidence validity and answer uniqueness. We evaluate 12 frontier models under a unified tool environment and execution protocol. The best-performing model achieves 39.0\% accuracy, while human evaluation across 1,000 independent attempts yields 21.1\% accuracy. Further analyses of tool usage, search cost, termination behavior, and failure recovery reveal substantial differences in research strategies across models, as well as persistent challenges in tool coordination and long-horizon search management. OpenDS-World provides a unified evaluation platform for systematically assessing and advancing agents' autonomous deep search capabilities in real-world open Web environments.

\end{abstract}%
}{}


\newpage

{%
  \hypersetup{linkcolor=rlTocLink}
  \tableofcontents
\thispagestyle{fancy}
}


\newpage

\ifdefstring{\abstractposition}{intro}{%
  \begin{abstract}

Despite substantial progress in Deep Search agents, their ability to conduct end-to-end research in real-world open Web environments remains insufficiently evaluated. Unlike static retrieval settings, information on the real-world Web is distributed across diverse sources, modalities, and heterogeneous formats, often involving complex interaction mechanisms that require agents to autonomously search for, access, and integrate relevant evidence during execution. To address this gap, we introduce the \textbf{OpenDS} task, designed to evaluate agents' ability to perform multi-step deep search in the live, open Web. This task requires agents to autonomously explore the real Web, discover and connect relevant evidence from multiple independent sources through Web search, browser interaction, and evidence processing, and ultimately determine a unique factual answer. Building on this task formulation, we construct \textbf{OpenDS-World}, a benchmark comprising 200 challenging tasks spanning diverse information modalities, heterogeneous file formats, and interactive access requirements, with independent human verification to ensure evidence validity and answer uniqueness. We evaluate 12 frontier models under a unified tool environment and execution protocol. The best-performing model achieves 39.0\% accuracy, while human evaluation across 1,000 independent attempts yields 21.1\% accuracy. Further analyses of tool usage, search cost, termination behavior, and failure recovery reveal substantial differences in research strategies across models, as well as persistent challenges in tool coordination and long-horizon search management. OpenDS-World provides a unified evaluation platform for systematically assessing and advancing agents' autonomous deep search capabilities in real-world open Web environments.

\end{abstract}%
}{}

\section{Introduction}

A central objective of Deep Search is to enable AI agents to autonomously acquire, analyze, and integrate information to solve complex information-seeking and research tasks~\citep{xi2025survey,zhang2025deep,zhang2025web}. Recent advances in large language models and tool-augmented agents have substantially improved their capabilities in multi-round search, Web browsing, and cross-source reasoning~\citep{jin2025hira, qiao2025webresearcher}, enabling them to address increasingly complex information needs~\citep{zheng2025deepresearcher,wu2026webdancer,li2025websailor, li2026webthinker}. Meanwhile, reliably evaluating these agents' research capabilities in practical environments has become an increasingly important issue. Existing studies have explored various aspects of agent capabilities, including general tool use, complex information retrieval, persistent Web search, and interactive Web navigation, providing an important foundation for understanding their search, reasoning, and tool-use capabilities~\citep{wei2025browsecomp,zhao2026lohosearch,wu2025webwalker, xue2025illusion}.

Despite substantial progress in open Web search, multi-round information retrieval, and cross-source reasoning, existing Deep Search evaluations remain insufficient for systematically assessing the complex evidence acquisition requirements of real-world Web environments. The complexity of the real-world Web is primarily reflected in three aspects: Diverse Modalities, where information is presented in various forms, including text, images, audio, and video~\citep{li2025mm, liu2026watching}; Heterogeneous Formats, where information is distributed across resources with different structures and processing requirements~\citep{jin2025hira}, such as HTML, PDF, XML, and Parquet; and Interactive Access Requirements, where certain evidence cannot be obtained through direct retrieval but requires interactive operations such as Web navigation, interface manipulation, parameter updates, or artifact transfer~\citep{liu2026uis, xu2026cap}. Together, these attributes constitute the complex evidence acquisition process in real-world Web environments, requiring agents to coordinate information discovery, interactive access, resource processing, and cross-source reasoning throughout multi-step research. However, existing benchmarks~\citep{li2025mm,java2026characterizing,liu2026uis,xu2026cap} have not sufficiently covered these interconnected challenges, limiting the systematic analysis of agents' deep search capabilities in real-world environments.

\begin{figure}[!t]
    \centering
    \includegraphics[width=1.0\textwidth]{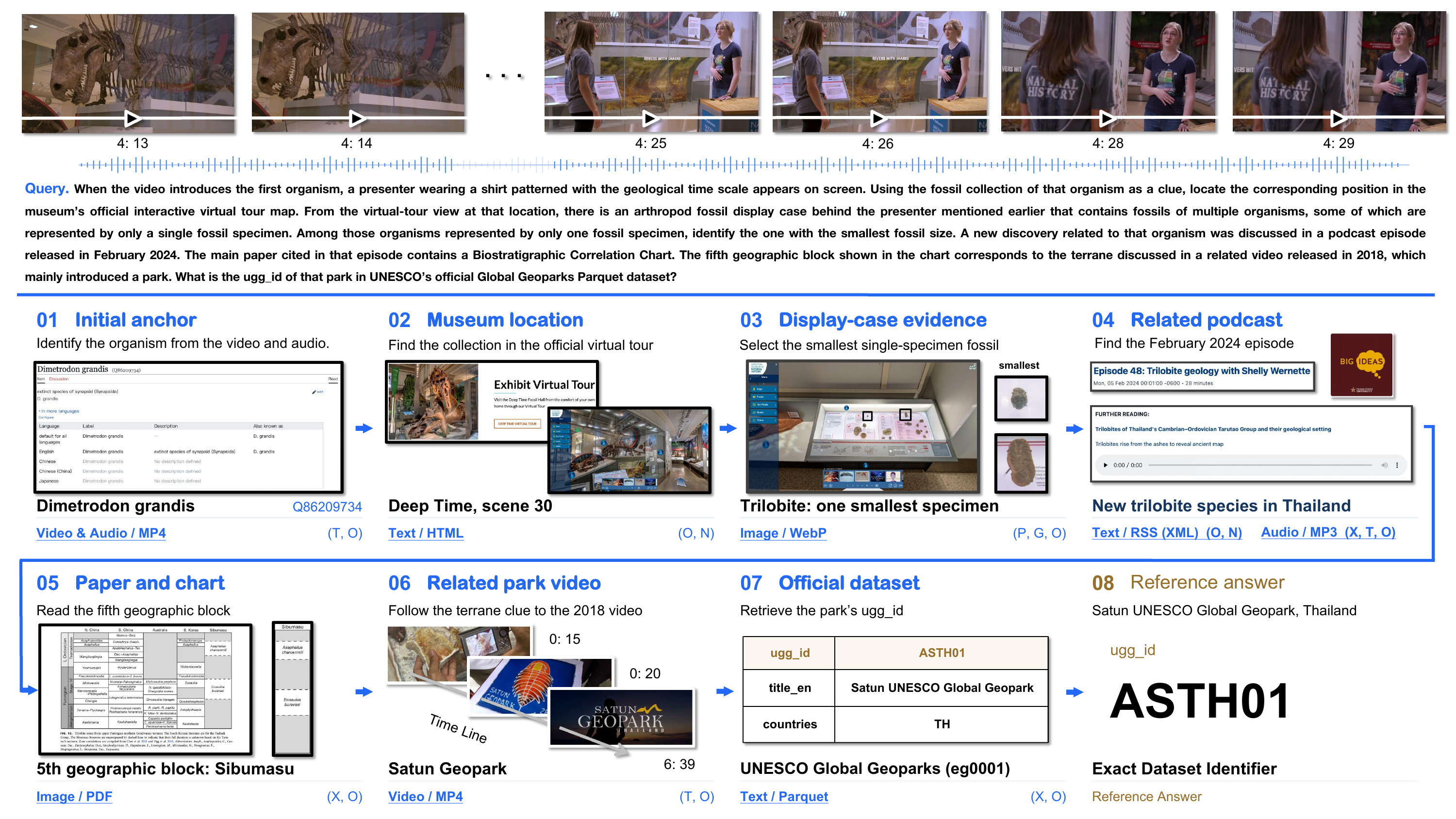}
    \caption{\textbf{An example of deep search in OpenDS-World.} The example connects audiovisual identification, virtual-tour navigation, document inspection, and structured-data retrieval across multiple Web sources. }
    \label{fig:intro}
    
\end{figure}

To address this gap, we introduce the \textbf{OpenDS (Open-Web Deep Search) task}, designed to systematically evaluate agents' ability to conduct deep search across heterogeneous information sources in real-world Web environments. This task requires agents to autonomously search, access, and connect evidence across Web sources with diverse information modalities, resource formats, and interactive access requirements to complete complex multi-step searches. Building upon the OpenDS task, we construct \textbf{OpenDS-World}, a benchmark comprising 200 challenging tasks designed around real-world Web information sources, spanning eight domains: culture, daily life, science, economy, history, nature, society, and technology. Each task requires agents to start from a video containing real-world information and independently discover evidence from multiple sources across the open Web to support the answer. As illustrated in Figure~\ref{fig:intro}, solving a task may involve accessing online videos, interactive virtual tours, research papers, and structured datasets, and acquiring and connecting relevant evidence through different access mechanisms. To systematically characterize the challenges of evidence acquisition in real-world Web environments, OpenDS-World covers diverse information modalities, heterogeneous file formats, and seven types of access requirements. We employ independent human verification and task filtering to ensure evidence validity, factual accuracy, and answer uniqueness, and further conduct human testing to assess practical research difficulty.

We systematically evaluate 12 frontier models under a unified tool environment and execution protocol. The results reveal that current agents still face substantial challenges when conducting deep search in real-world Web environments. The best-performing model, GPT-6 Astra~\citep{openai2026gpt6astra}, achieves only 39.0\% accuracy across all 200 tasks. By comparison, human participants achieve an overall accuracy of 21.1\% in independent solving attempts (Figure~\ref{fig:opends}), with 42.5\% of tasks remaining unsolved by all assigned human participants. Further analysis reveals substantial differences across models in tool usage, search cost, termination behavior, and failure recovery. More tool calls and longer execution times do not necessarily lead to higher accuracy, reflecting limitations in current agents' information-source selection, tool coordination, search strategy adjustment, and long-horizon execution management.


By covering diverse information modalities, heterogeneous resource formats, interactive access requirements, and multi-step evidence dependencies, OpenDS-World provides a unified platform for systematically evaluating agents' ability to conduct deep search across heterogeneous information sources in real-world Web environments. Our main contributions are summarized as follows:


\vspace{-5pt}

\begin{itemize}
    \item \textbf{Deep Search Task in Real-World Web Environments (OpenDS Task).} We introduce the OpenDS task to systematically evaluate agents' ability to conduct deep search across heterogeneous information sources in real-world Web environments. The task requires agents to autonomously search, access, process, and connect evidence across information sources with diverse modalities, resource formats, and interactive access requirements, determining a unique factual answer through multi-step research.

    \item \textbf{The OpenDS-World Benchmark.} Building upon the OpenDS task, we construct a benchmark comprising 200 challenging tasks across eight domains, covering diverse information modalities, heterogeneous resource formats, and seven types of access requirements.

    \item \textbf{Systematic Model Evaluation.} We evaluate 12 frontier models under a unified tool environment and execution protocol, systematically analyzing task accuracy, search cost, tool usage, termination behavior, and failure recovery. Our analysis reveals differences in search strategies and execution behaviors across models, as well as key challenges faced by current agents when conducting deep search in complex real-world Web environments.
\end{itemize}


\newpage

\begin{figure}[!t]
    \centering
        \includegraphics[width=1.0\textwidth]{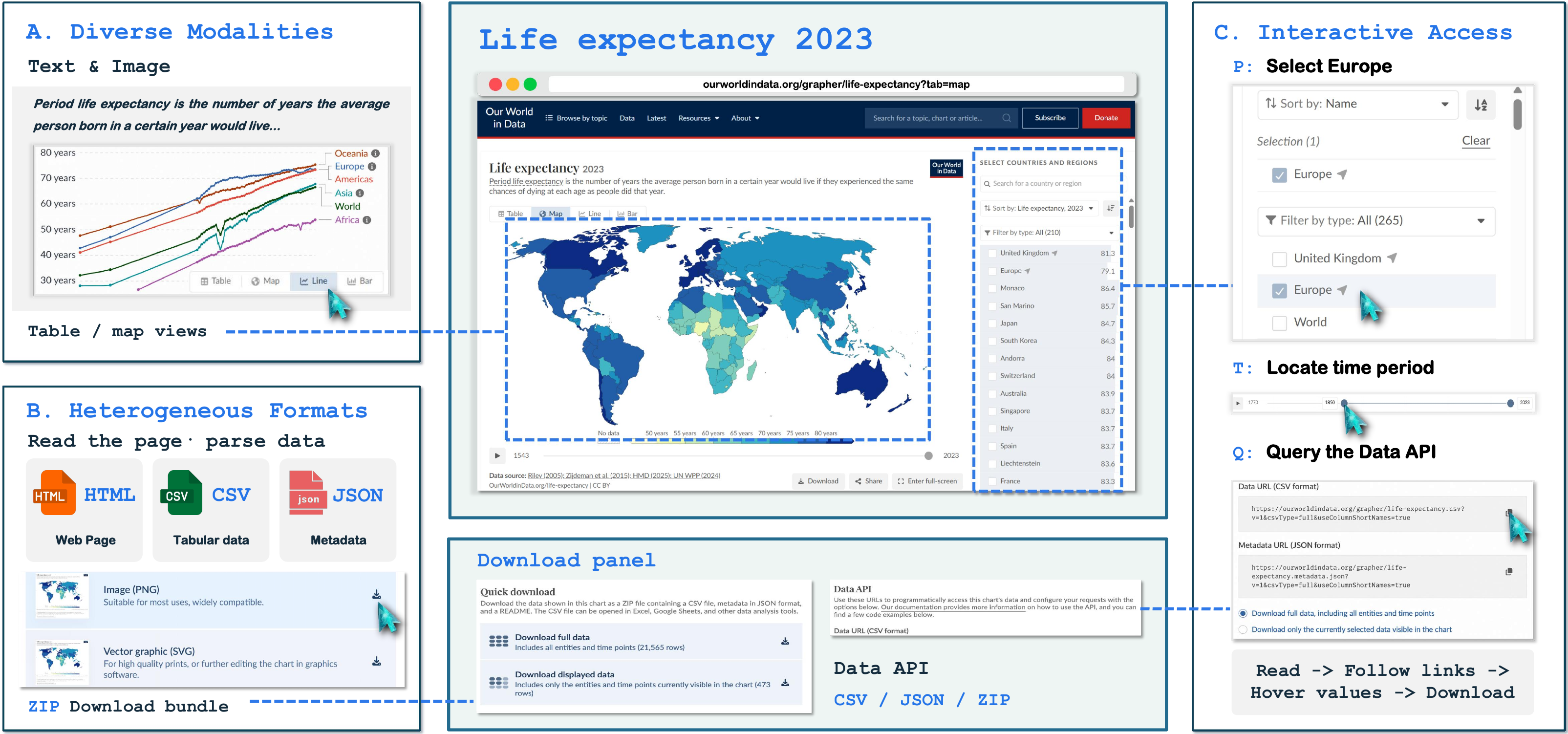}
    \caption{\textbf{Three complementary properties of a real-world Web source.}
    An Our World in Data page illustrates how a single source combines (A) diverse modalities, including text and visual charts; (B) heterogeneous formats, including HTML, CSV, JSON, and downloadable ZIP bundles; and (C) interactive access, including region selection, temporal indexing, and declarative API queries. Obtaining relevant evidence requires navigating these connected representations and source states.}
    \label{fig:example1}
\end{figure}

\section{OpenDS-World}

We introduce \textbf{OpenDS-World}, a benchmark for systematically evaluating AI agents' ability to conduct deep search across heterogeneous information sources in real-world Web environments. The benchmark comprises 200 challenging tasks constructed from real-world Web information sources, spanning eight domains: culture, daily life, science, economy, history, nature, society, and technology (Figure~\ref{fig:app-S01_dataset}(a)). OpenDS-World focuses on the complex process of evidence acquisition in real-world Web environments, evaluating agents' ability to autonomously discover, acquire, and connect evidence across different information sources involving diverse information modalities, heterogeneous resource formats, and interactive access requirements.

\subsection{Open-Web Deep Search}

\subsubsection{Task Definition}

We define \textbf{Open-Web Deep Search (OpenDS)} as a factual question-answering task that involves multi-step evidence search and reasoning across heterogeneous information sources in real-world open Web environments. Given a factual question and available initial clues, an agent must autonomously explore the open Web, continuously search for, access, process, and connect relevant evidence across information sources with diverse modalities, resource formats, and interactive access requirements, ultimately determining a unique and verifiable factual answer. In OpenDS, the evidence required to complete a task may be distributed across different websites, documents, multimedia resources, and online applications, while its specific locations and access methods are not provided to the agent in advance. The agent must determine subsequent search targets based on the information currently available, select appropriate information access and processing methods, and progressively establish connections among evidence through cross-source reasoning. Because different information sources vary in content representation, data organization, and access mechanisms, the agent must flexibly coordinate search, interaction, and information processing operations throughout multi-step research, enabling evidence from different sources to be connected and collectively support the final answer. Beyond information discovery and cross-source reasoning~\citep{li2025search, li2026webthinker}, OpenDS emphasizes the continuous process of evidence acquisition and integration across different information modalities, resource formats, and access mechanisms in real-world Web environments. These interconnected requirements collectively constitute the core task characteristics of OpenDS.


\subsubsection{Main Attributes}

OpenDS is organized around three main attributes of real-world Web information environments: \textbf{Diverse Modalities}, \textbf{Heterogeneous Formats}, and \textbf{Interactive Access Requirements}. These attributes characterize the complexity of the real-world Web in terms of information representation, organizational structure, and access methods, respectively. They jointly influence how agents discover, acquire, and connect evidence across different information sources throughout the search process. Figure~\ref{fig:example1} illustrates how these three attributes coexist within a single Web source.


\textbf{Diverse Modalities.} Information on the real-world Web is represented across multiple modalities, including text, images, audio, and video. Relevant evidence may appear in Web text, visual content, spoken information, or multimedia resources, and different stages of the same research task may depend on information from different modalities. Therefore, agents need to continuously search across information sources of different modalities, interpret the clues provided by various types of content, and establish cross-modal evidence connections~\citep{li2025mm,tian2025mmina,liu2026watching}. OpenDS-World evaluates agents' capabilities in cross-modal search and evidence integration through evidence paths involving multiple information modalities.

\textbf{Heterogeneous Formats.} Information on the real-world Web is organized in different file and data formats, such as HTML pages, PDF documents, WebP images, MP3 audio, MP4 video, RSS/XML data sources, and structured data formats such as Parquet. Different formats have distinct content structures, parsing methods, and processing requirements~\citep{liu2026uis, jin2025hira, yuan2026osworld2}. Even when information belongs to the same modality, its acquisition and processing methods may differ substantially. During cross-source search, agents need to select appropriate access and processing methods according to resource formats, extract relevant evidence from heterogeneous data, and use the acquired information for subsequent search and reasoning. OpenDS-World evaluates agents' ability to conduct continuous search and evidence integration across heterogeneous resources through combinations of information sources with different formats.

\textbf{Interactive Access Requirements.} Some information on the real-world Web cannot be obtained through direct observation or simple retrieval. Instead, obtaining it requires specific access operations to reveal or extract the relevant evidence~\citep{liu2026uis,xue2025illusion,xu2026cap}. These operations may include page navigation, parameter input, temporal indexing, graphical interface manipulation, structured queries, and resource downloads. For example, locating a particular exhibit in an interactive museum virtual tour may require an agent to autonomously explore the interface, switch viewpoints, and identify relevant visual evidence. Access requirements may vary substantially across different information sources. Therefore, agents need to dynamically select operations according to the location and access mechanisms of the target evidence, and establish connections between the acquired information and evidence from other sources. OpenDS-World incorporates different access requirements into task evidence paths to evaluate agents' ability to continuously acquire and connect information under complex interactive conditions.



\subsection{OpenDS-World Construction}

\subsubsection{Task Construction}

\textbf{Stage 1. Real-World Anchor.} Each evidence path begins with a publicly accessible video. Expert annotators select videos spanning eight domains: culture, daily life, science, economy, history, nature, society, and technology. In total, 500 candidate videos are selected for task construction. They jointly analyze consecutive visual frames and audio content to identify a verifiable initial anchor. The anchor corresponds to an entity whose identification must rely on the audiovisual content of the video.
This constraint prevents anchor identification from degenerating into page-level textual search~\citep{li2025mm, liu2026watching}. To ensure stable access and reproducibility, the videos used during task construction are archived and distributed with the benchmark as part of the task input, with each video matched to its corresponding task by ID.
To ensure an unambiguous semantic reference, the core entity associated with the anchor must uniquely correspond to a valid Wikidata entry. The video, initial anchor, and corresponding Wikidata entry serve as the starting point for subsequent evidence expansion.

\textbf{Stage 2. Open-Web Evidence Expansion.} At each expansion step along the evidence path, expert annotators first construct a search query based on the current node, and then use Google Search to obtain a candidate set of Web information sources. Annotators inspect the candidate results one by one and select sources while ensuring that each source has a clear semantic relation to the current node and that the evidence is verifiable~\citep{huang2026gta}. During candidate source selection, annotators also consider the modality, format, and access-type attributes to increase the diversity of the final dataset across different forms of Web information. Annotators then extract a new entity from the selected source and designate it as the next node. Although each new node is initially identified from a selected source, its validity is not tied to that source alone. Each node along the evidence path must be independently corroborated by multiple Web sources, ensuring multi-source support throughout the entire evidence chain. This process is recursively applied along the evidence path~\citep{tao2026webshaper} until reaching a unique and stable terminal node suitable as the reference answer. In this way, the evidence path is dynamically constructed through iterative search, access, and verification over real Web information sources.

\textbf{Stage 3. Question Inversion.} Once the evidence path reaches a unique terminal node, we designate it as the reference answer and construct the textual question by inverting the complete evidence path into a set of interdependent constraints~\citep{wei2025browsecomp,zhao2026lohosearch}. The intermediate nodes are not explicitly revealed. Instead, each node is transformed into an indirect referring description derived from the evidence accumulated along the preceding path~\citep{lu2025deepdive, liu2025webexplorer}.
These descriptions are composed so that no individual clue directly exposes the target entity, while their combination progressively constrains the candidate space and uniquely determines the reference answer. As a result, directly retrieving the answer from an explicit clue cannot solve the question. It requires recovering the latent correspondences among intermediate entities and following the underlying evidence chain.

\subsubsection{Human Verification}

We conduct independent human verification for each candidate task, involving a total of ten independent reviewers. Reviewers revisit the complete evidence path using the source links preserved during annotation, including the initial anchor, intermediate entities, inter-entity relations, and the final reference answer. The verification process consists of two main aspects: evidence validity and task validity.

\textbf{Evidence Validity.} Reviewers examine the entities along the evidence path, their semantic relations, and the final reference answer to ensure factual accuracy and support from authentic, accessible, and independently verifiable Web evidence~\citep{chen2025browsecomp, xu2026cap}. To avoid over-reliance on the original source, reviewers treat it as just one reference among several and cross-check the information against multiple additional independent Web sources.
They further verify that each transition between adjacent nodes is explicitly supported by evidence, ensuring factual and logical consistency throughout the evidence path. For the initial Real-world Anchor, reviewers additionally confirm that it can be uniquely identified from the visual and audio content of the video, that its core entity is correctly and uniquely mapped to a valid Wikidata entry, and that the relevant information cannot be directly obtained from the textual content on the same video page.

\textbf{Task Validity.} Reviewers further examine whether each candidate task satisfies the construction principles, focusing on question clarity, constraint consistency, and answer uniqueness. Specifically, reviewers verify that the key clues in the question accurately correspond to the relevant information along the evidence path, that each constraint is supported by verifiable evidence~\citep{java2026characterizing}, and that no factual or logical conflicts exist among the constraints. Reviewers also confirm that these constraints collectively determine a unique factual answer, avoiding ambiguity arising from insufficient clues, unclear entity references, or multiple plausible answers~\citep{wei2025browsecomp, li2025mm}. Any candidate task containing factual errors, insufficient evidence, inconsistent relations, ambiguous questions, non-unique reference answers, or violations of the construction principles is directly excluded.

Following the above filtering procedures, we construct the final OpenDS-World benchmark comprising 200 tasks. All tasks have a clearly defined reference answer and a verifiable evidence path.

\subsection{Human Testing}

After completing human filtering, we further conduct an independent human evaluation to assess the practical difficulty of solving the questions in a real Web environment. This evaluation is to measure how difficult it is for independent solvers to discover the evidence path and to provide human performance data for subsequent difficulty calibration.

\begin{figure}[!t]
    \centering
    \includegraphics[width=1.0\textwidth]{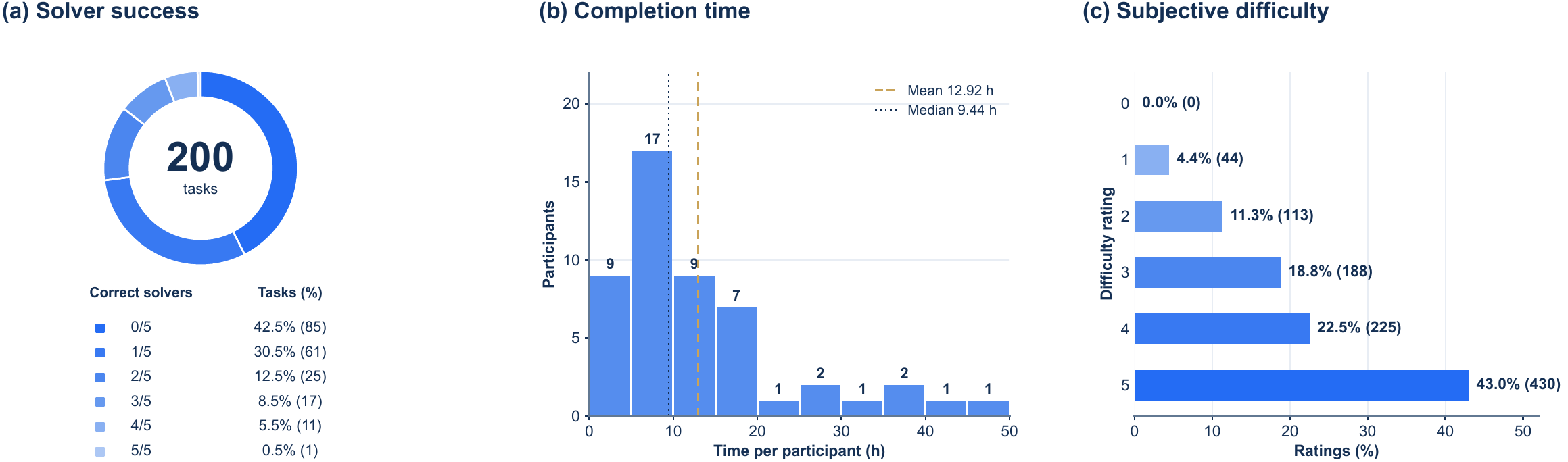}
    \caption{\textbf{Human performance, solving time, and perceived difficulty on OpenDS-World.}
    (a) Distribution of the 200 tasks by the number of correct solvers among five independent participants per task.
    (b) Total solving time for each of the 50 participants across their assigned set of 20 tasks; vertical lines indicate the mean and median.
    (c) Distribution of 1,000 subjective difficulty ratings on a six-point scale from 0 (easiest) to 5 (most difficult).
    Bar annotations report percentages and counts in (a) and (c), and participant counts in (b).}
    \label{fig:solver}
\end{figure}

\subsubsection{Independent Human Solving}

We divide the 200 questions into 10 disjoint sets of 20 questions, with each set independently attempted by a group of five human solvers, for a total of 50 participants. Each solver completes only one question set to control individual workload and reduce potential fatigue effects. None of the solvers participated in constructing or filtering the questions they answered, and communication among solvers was not allowed during testing. Human solvers are given only the complete original input, including the video, audio, and textual question. They are not provided with the reference answer or the evidence path constructed during annotation. Solvers may use standard Web browsers such as Google Chrome for conventional Web browsing and search, but they are not allowed to use any LLM-based systems or tools. No uniform time limit is imposed for individual questions. The solving time for each question is measured from the moment a solver begins viewing the task input until the solver voluntarily terminates the attempt. When a solver considers that no further effective progress can be made, the solver may terminate the current question. Each human solver is required to submit a final answer, the total time spent solving the question, and a six-point subjective difficulty rating from 0 to 5, where 0 indicates the easiest and 5 indicates the most difficult. All final answers submitted by the solvers are manually checked to determine whether they are correct. During human testing, if any issue that affects normal solving is identified, the corresponding instance is recorded and jointly reviewed and revised after the testing is completed.

\subsubsection{Human Performance Analysis}

\textbf{Human Solvability.} The results of independent human solving indicate that OpenDS-World is highly challenging for human solvers. Across all 1,000 independent attempts, 211 resulted in correct answers, corresponding to an overall human accuracy of 21.10\%. At the task level, 115 (57.50\%) of the 200 tasks were successfully solved by at least one of the five independent solvers, while the remaining 85 (42.50\%) were not solved by any solver. Examining the number of successful solvers per task further shows that only 1 task (0.50\%) was successfully solved by all five solvers, while 61 (30.50\%), 25 (12.50\%), 17 (8.50\%), and 11 (5.50\%) tasks were successfully solved by one, two, three, and four solvers, respectively. In other words, only 29 tasks (14.50\%) were successfully completed by a majority of the solvers (at least three) (Figure~\ref{fig:solver}(a)). This distribution shows that a substantial portion of OpenDS-World remains highly challenging even for human solvers with unrestricted access to conventional Web browsers, while more than half of the tasks can still be successfully solved by at least one independent solver through actual Web research.

\textbf{Human Solving Effort.} This difficulty is also reflected in the research effort required to complete the tasks. As shown in Figure~\ref{fig:solver}(b), the 50 participants spent an average of 12.92 hours completing their assigned sets of 20 tasks, with a median of 9.44 hours and a combined total of approximately 646 hours. Across all 1,000 independent attempts, participants spent approximately 38.8 minutes per attempt on average. Meanwhile, across all 1,000 subjective difficulty ratings, the mean rating was 3.884/5, with a median of 4/5. Among these ratings, 43.00\% of the attempts received the highest difficulty rating of 5, another 22.50\% received a rating of 4, and no attempt received a rating of 0. Overall, 65.50\% of the human attempts were rated as difficulty level 4 or 5 (Figure~\ref{fig:solver}(c)). The human testing results show that OpenDS-World involves substantial practical research difficulty.

\begin{figure}[!t]
    \centering
    \includegraphics[width=1.0\textwidth]{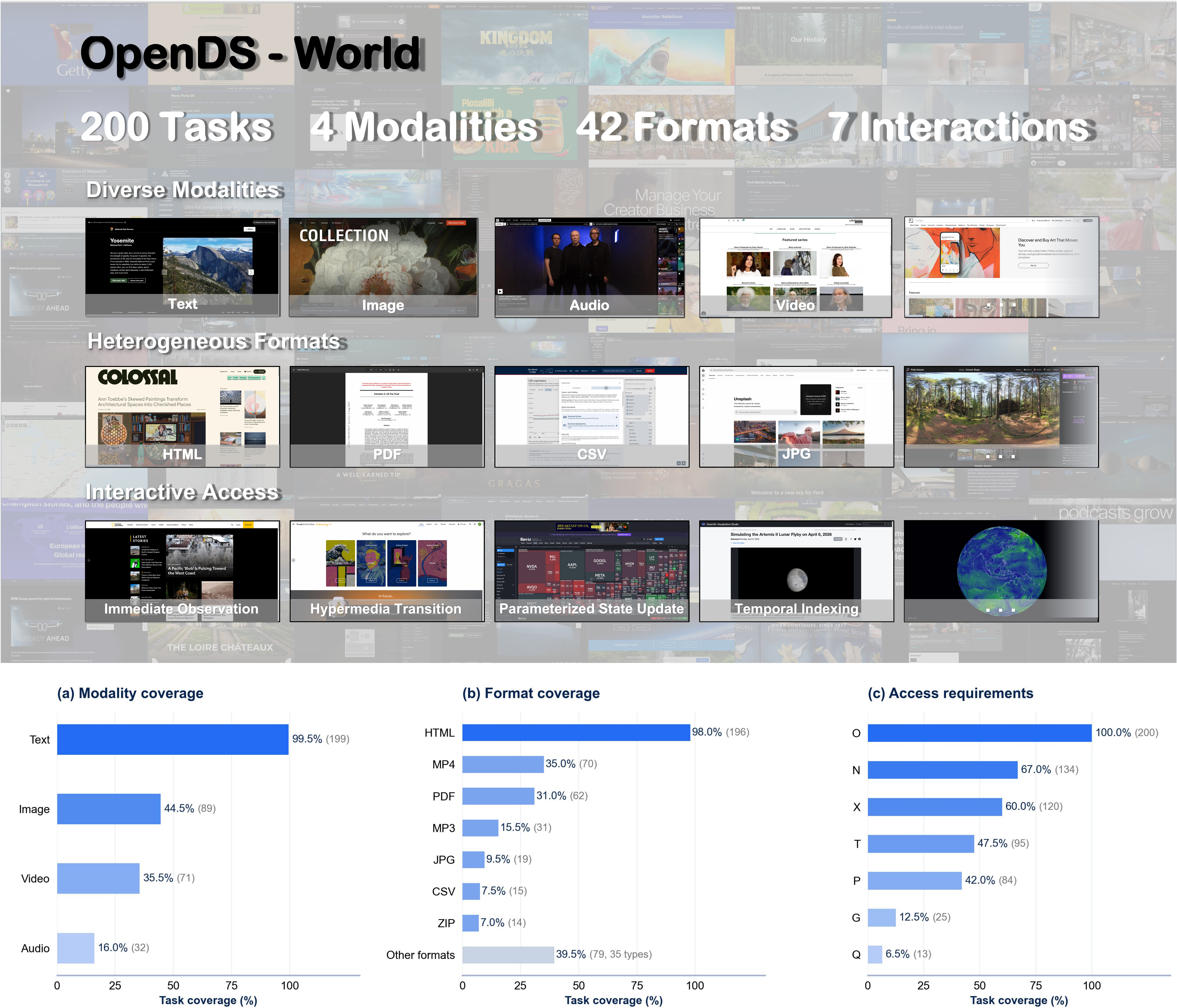}
    \caption{\textbf{Task diversity and statistics of OpenDS-World.}
    Each annotated evidence-chain entry is characterized by its modality, resource format, and access type, collectively covering four modalities, 42 formats, and seven access types.
    Top: representative Web sources illustrating these three dimensions.
    Bottom: (a) modality coverage across 200 tasks, (b) format coverage across 200 tasks, and (c) task coverage of interactive access types. Access-type codes in (c) are defined in Table~\ref{tab:access-taxonomy}.
    Parentheses report counts; Other formats comprise 108 entries spanning 35 types.
    Modality, format and access categories can overlap within a task.}
    \label{fig:example}
\end{figure}

\subsection{Task Statistics}

OpenDS-World contains 200 tasks, with 787 annotated entries for Web pages and other online resources along their evidence chains. These entries average 3.94 per task, and their task-level count distribution is shown in Figure~\ref{fig:app-S01_dataset}(c). Each entry is characterized by three attributes: the modality in which evidence is presented, the format of the resource carrying it, and the access type used to obtain it. Together, these annotations cover four modalities, 42 formats, and seven access types, describing both the information encountered during deep search and the interactions required to reach it. At the task level, text appears in 99.5\% of tasks, yet the evidence extends well beyond textual content. Images and video occur in 44.5\% and 35.5\% of tasks, respectively, while audio appears in 16.0\% (Figure~\ref{fig:example}(a)). Within individual tasks, 82.5\% involve at least two evidence modalities (Figure~\ref{fig:app-S02_diversity}(a)). This diversity is also reflected in resource formats. At the task level, HTML covers 98\% of tasks, while MP4 and PDF account for 35\% and 31\% (Figure~\ref{fig:example}(b)), respectively. At the task level, 96.0\% involve at least two resource formats
(Figure~\ref{fig:app-S02_diversity}(b)). At the annotated-entry level, HTML accounts for 56.5\% of entries, and MP4 and PDF together contribute 19.3\%. The remaining entries span audio, images, structured data, archives, and a long tail of less common formats (Table~\ref{tab:app-long-tail-formats}). Access requirements further reveal the interactive nature of the evidence chains. Although every task includes directly observable evidence, 67.0\% also involve navigating links, pages, or nested content, and 60.0\% involve transferring artifacts for further inspection. Temporal indexing occurs in 47.5\% of tasks, reflecting the need to locate evidence within time-based media, while 42.0\% involve parameterized state updates (Figure~\ref{fig:example}(c)), such as applying filters or selecting a year or region. Graphical manipulation and declarative queries extend this coverage to visual interfaces and structured information access. Multiple access types occur in 95.5\% of tasks (Figure~\ref{fig:app-S02_diversity}(c)). These complementary statistics show how deep search unfolds across heterogeneous online resources, requiring evidence discovery through multiple forms of content and interaction. 

\clearpage

\section{Experiments}

\subsection{Setting}

\paragraph{Models.} We evaluate 12 frontier models: GPT-6 Astra~\citep{openai2026gpt6astra}, GPT-5.6 Sol~\citep{openai2026gpt56}, Claude Opus 5.5~\citep{anthropic2026opus55}, Claude Fable 5.1~\citep{anthropic2026fable51}, Gemini 3.8 Flash~\citep{googledeepmind2026gemini38flash}, Qwen3.8-Max~\citep{qwen38}, Qwen3.8-Flash~\citep{qiu2026design}, DeepSeek-V4.1-Flash~\citep{deepseekai2026deepseekv41flash}, DeepSeek-V4-Pro~\citep{deepseekai2026deepseekv4}, Kimi K3~\citep{team2026kimi}, GLM-5.3-Flash~\citep{glm5team2026glm5vibecodingagentic}, and GLM-5.3~\citep{glm5team2026glm5vibecodingagentic}. All models are evaluated using the same task inputs, system prompts, tool environment, and execution budgets to ensure comparability across models. The research and finalization prompts are provided in Appendix~\ref{app:agent-prompts}.

\paragraph{Harness.} Given the characteristics of this task, we build a unified agent harness that provides models with three core tool capabilities: \textbf{Search Tool} supports text, image, and video search for locating potential information sources; \textbf{GUI Tool} supports web navigation, content reading, clicking, input, scrolling, and screenshots for accessing and interacting with dynamic web pages; and \textbf{Other Tools} support video-frame inspection and subtitle extraction for both input and external videos. Table~\ref{tab:tool} lists the tool interfaces, arguments, and supported operations. To ensure that different models are evaluated under consistent computational and tool-use budgets, we apply the same execution constraints to every sample. The search phase for each sample is limited to 3,600 seconds, and all tools share a maximum budget of 100 tool calls. Further search is terminated once either limit is reached. Additional execution and context-management settings are summarized in Table~\ref{tab:agent-execution-settings}. The task ends when the model successfully generates a final answer. If the search phase terminates due to timeout, exhaustion of the tool budget, or failure to generate a valid answer, the system enters a tool-disabled finalization stage and generates the final answer solely based on the evidence collected up to that point.

\paragraph{Evaluation.} We use Qwen3.8-27B~\citep{qwen38} as the LLM judge. The judge receives the reference answer and the model answer and determines whether the model answer is correct based on their semantic equivalence~\citep{wei2025browsecomp, liu2026watching}. We use answer accuracy (ACC) as the primary metric and report total token consumption and elapsed time as resource-cost metrics. The judging protocol and prompts are provided in Appendix~\ref{app:judge-evaluationllm}.

\begin{table}[t]
\centering
\caption{\textbf{Main results on OpenDS-World.}
We evaluate frontier models across three difficulty levels which are defined by the number of correct human solvers out of the five participants assigned to each task: Low ($\geq 2$; 54 tasks), Medium ($1$; 61 tasks), and High ($0$; 85 tasks). Evidence-attribute coverage within these difficulty groups is reported in Figure~\ref{fig:app-S03_attributes_by_difficulty} (Appendix~\ref{sec:source-requirements-difficulty}).
For each level, we report accuracy (\textbf{Acc}, \%),
total wall-clock time (\faIcon[regular]{clock} / h),
and total token consumption (\textbf{Token / M.}).
The shaded \textbf{Overall} columns summarize all 200 tasks.
Numbers in parentheses indicate correct answers. Bold accuracy values denote the best result in each group. The superscript \textsuperscript{(T)} denotes text-only models.}
\label{tab:main-results}

\begingroup
\small
\renewcommand{\arraystretch}{1.26}
\setlength{\tabcolsep}{4.2pt}
\setlength{\aboverulesep}{3pt}
\setlength{\belowrulesep}{3pt}
\setlength{\heavyrulewidth}{0.8pt}
\setlength{\lightrulewidth}{0.4pt}

\resizebox{\textwidth}{!}{%
\begin{tabular}{
@{}l
rrr
@{\hspace{12pt}}rrr
@{\hspace{12pt}}rrr
@{\hspace{10pt}}
>{\columncolor{odOverall}}r
>{\columncolor{odOverall}}r
>{\columncolor{odOverall}}r
@{\hspace{5pt}}
}
\toprule

& \multicolumn{3}{c}{\textsf{\textbf{Low}}}
& \multicolumn{3}{c}{\textsf{\textbf{Medium}}}
& \multicolumn{3}{c}{%
\textcolor{odAccent}{\textsf{\textbf{High}}}}
& \multicolumn{3}{>{\columncolor{odOverall}}c}{%
\textsf{\textbf{Overall}}}
\\[-2pt]

& \multicolumn{3}{c}{\odtasks{54}}
& \multicolumn{3}{c}{\odtasks{61}}
& \multicolumn{3}{c}{\odtasks{85}}
& \multicolumn{3}{>{\columncolor{odOverall}}c}{%
\odtasks{200}}
\\

\cmidrule(lr){2-4}
\cmidrule(lr){5-7}
\cmidrule(lr){8-10}
\cmidrule(lr){11-13}

\textbf{Model}
& Acc (\%) & \faIcon[regular]{clock} & Token.
& Acc (\%) & \faIcon[regular]{clock} & Token.
& Acc (\%) & \faIcon[regular]{clock} & Token.
& \textbf{Acc (\%)} & \faIcon[regular]{clock} & Token.
\\
\midrule

\modellogo{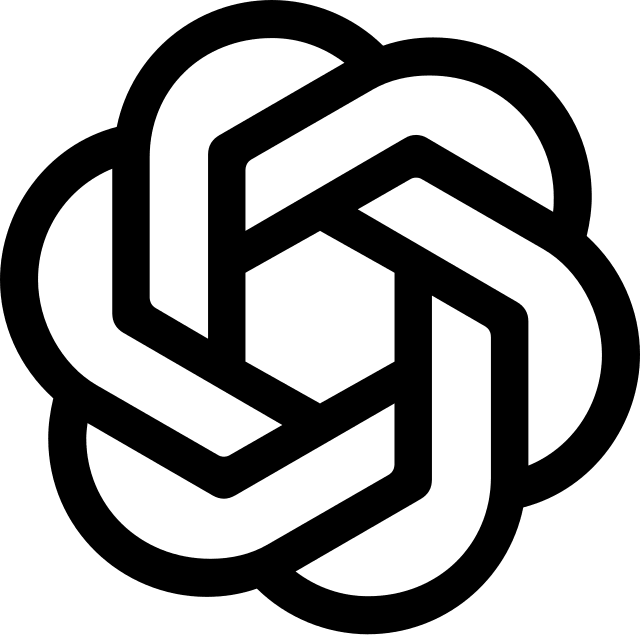}\hspace{0.35em}GPT-6 Astra
& \odbest{59.26}{32} & 13 & 36
& \odbest{40.98}{25} & 17 & 55
& \odacc{24.71}{21} & 32 & 96
& \odbest{39.0}{78} & 62 & 187
\\

\modellogo{openai.png}\hspace{0.35em}GPT-5.6 Sol
& \odacc{46.30}{25} & 6 & 34
& \odacc{31.15}{19} & 9 & 68
& \odacc{14.12}{12} & 15 & 84
& \odacc{28.0}{56} & 30 & 186
\\

\modellogo{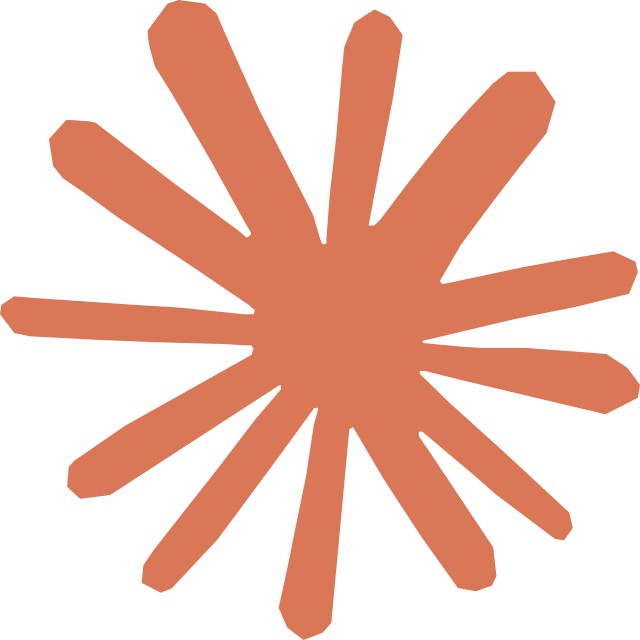}\hspace{0.35em}Claude Opus 5.5
& \odacc{53.70}{29} & 7 & 75
& \odacc{\textbf{40.98}}{25} & 9 & 112
& \odacc{25.88}{22} & 50 & 151
& \odacc{38.0}{76} & 65 & 339
\\

\modellogo{claude.png}\hspace{0.35em}Claude Fable 5.1
& \odacc{48.15}{26} & 24 & 99
& \odacc{32.79}{20} & 34 & 170
& \odacc{24.71}{21} & 52 & 212
& \odacc{33.5}{67} & 110 & 481
\\

\modellogo{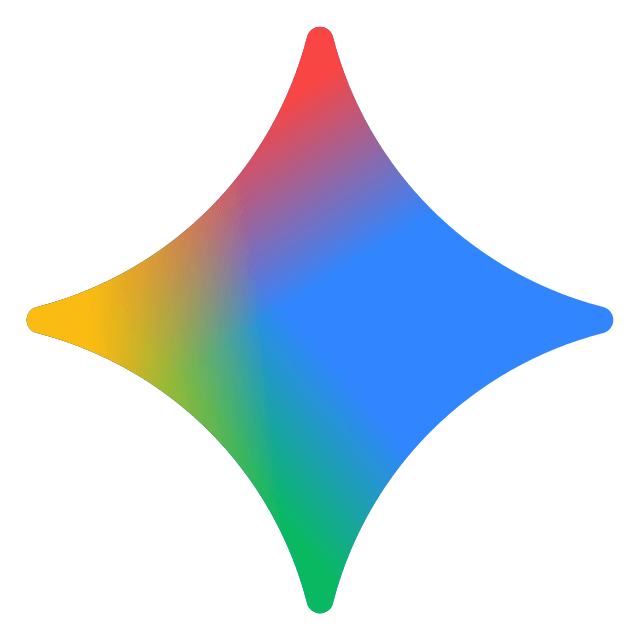}\hspace{0.35em}Gemini 3.8 Flash
& \odacc{51.85}{28} & 13 & 194
& \odacc{34.43}{21} & 20 & 373
& \odbest{28.24}{24} & 26 & 409
& \odacc{36.5}{73} & 59 & 976
\\

\modellogo{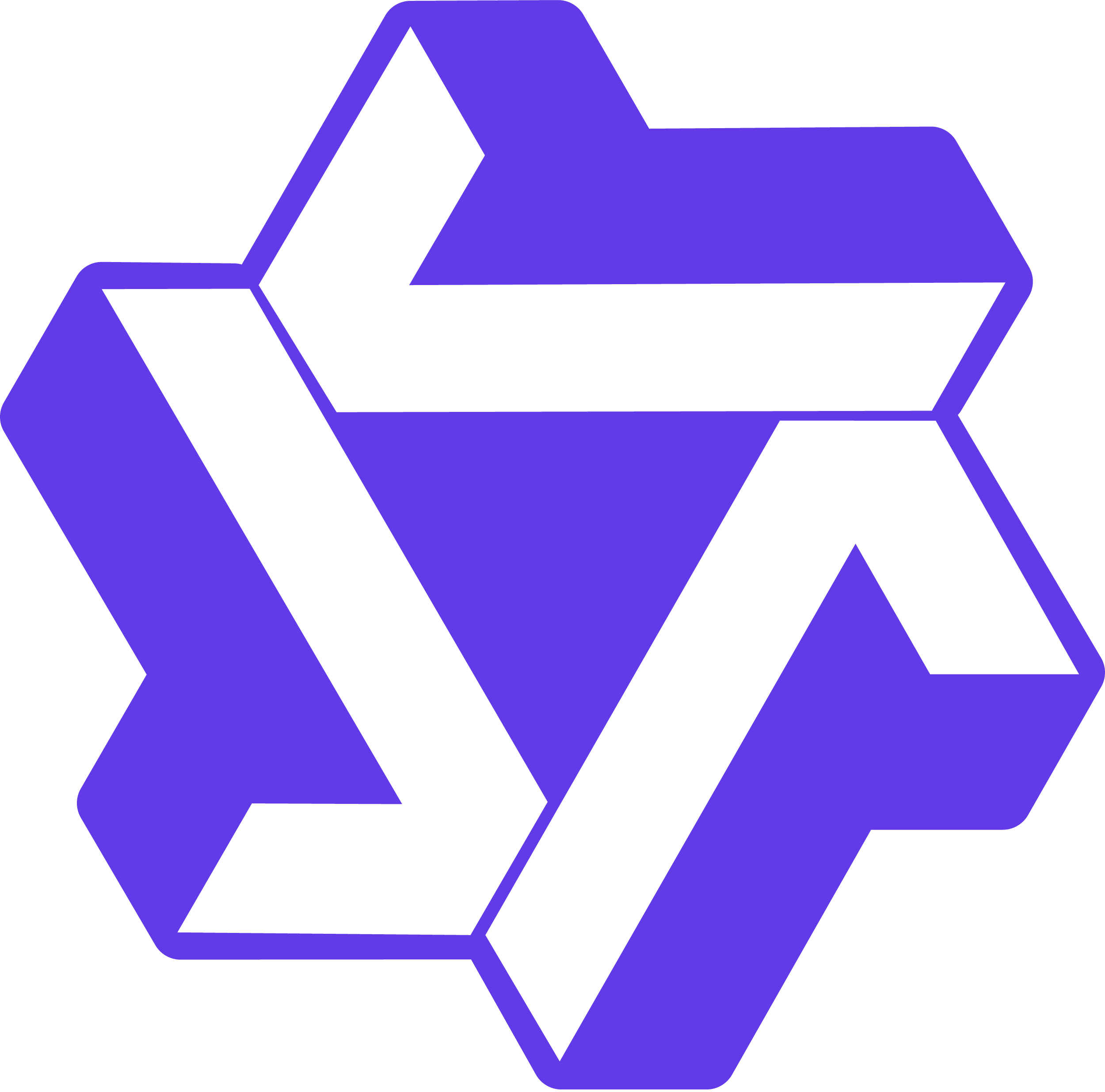}\hspace{0.35em}Qwen3.8-Max
& \odacc{40.74}{22} & 24 & 223
& \odacc{26.23}{16} & 28 & 256
& \odacc{15.29}{13} & 33 & 360
& \odacc{25.5}{51} & 85 & 839
\\

\modellogo{qwen.png}\hspace{0.35em}Qwen3.8-Flash
& \odacc{22.22}{12} & 20 & 187
& \odacc{13.11}{8} & 24 & 220
& \odacc{4.71}{4} & 32 & 345
& \odacc{12.0}{24} & 75 & 752
\\

\modellogo{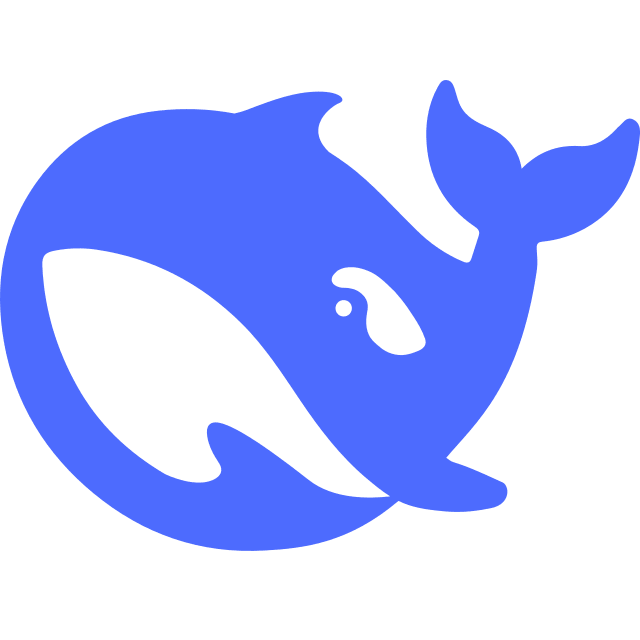}\hspace{0.35em}DeepSeek-V4.1-Flash
& \odacc{33.33}{18} & 31 & 141
& \odacc{13.11}{8} & 37 & 182
& \odacc{17.65}{15} & 52 & 276
& \odacc{20.5}{41} & 120 & 599
\\

\modellogo{deepseek.png}\hspace{0.35em}DeepSeek-V4-Pro
\textsuperscript{\textbf{(T)}}
& \odacc{27.78}{15} & 34 & 80
& \odacc{11.48}{7} & 49 & 101
& \odacc{3.53}{3} & 65 & 115
& \odacc{12.5}{25} & 148 & 297
\\

\modellogo{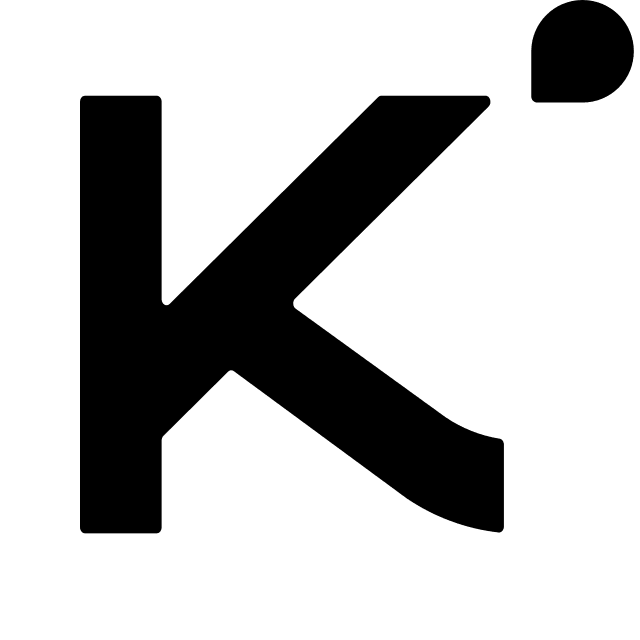}\hspace{0.35em}Kimi K3
& \odacc{27.78}{15} & 24 & 65
& \odacc{14.75}{9} & 33 & 104
& \odacc{12.94}{11} & 52 & 141
& \odacc{17.5}{35} & 110 & 310
\\

\modellogo{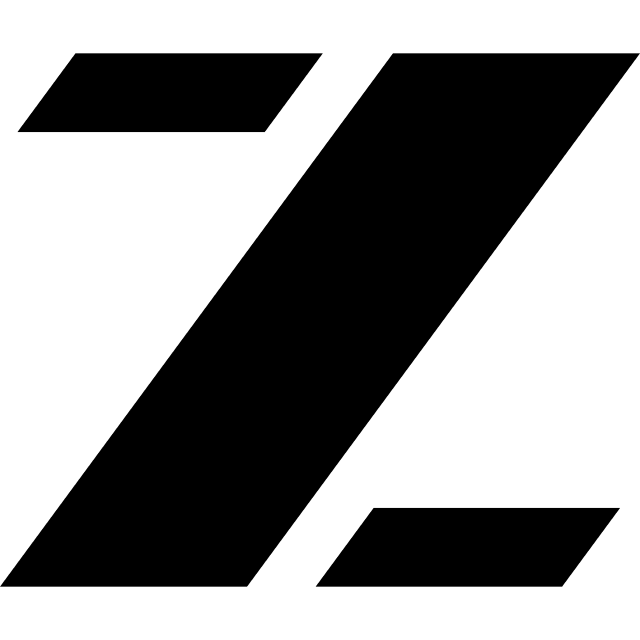}\hspace{0.35em}GLM-5.3-Flash
& \odacc{42.59}{23} & 29 & 67
& \odacc{16.39}{10} & 42 & 137
& \odacc{16.47}{14} & 56 & 186
& \odacc{23.5}{47} & 128 & 389
\\

\modellogo{zai.png}\hspace{0.35em}GLM-5.3
\textsuperscript{\textbf{(T)}}
& \odacc{22.22}{12} & 35 & 76
& \odacc{6.56}{4} & 45 & 114
& \odacc{12.94}{11} & 65 & 155
& \odacc{13.5}{27} & 145 & 345
\\

\bottomrule
\end{tabular}%
}
\endgroup
\end{table}

\subsection{Main Results}

Table~\ref{tab:main-results} presents the main results on OpenDS-World. Overall, current frontier models remain limited in performing Deep Search in real-world Web environments. GPT-6 Astra achieves the highest overall accuracy of 39.00\% over 200 tasks, followed by Claude Opus 5.5 at 38.00\%, Gemini 3.8 Flash at 36.50\%, and Claude Fable 5.1 at 33.50\%. GPT-5.6 Sol and Qwen3.8-Max achieve 28.00\% and 25.50\%, respectively, while GLM-5.3-Flash and DeepSeek-V4.1-Flash reach 23.50\% and 20.50\%. The remaining four models score below 20\%. Even the strongest model fails on 122 of the 200 tasks, corresponding to a failure rate of 61.00\%. These results highlight the persistent difficulty of finding and integrating information through deep search in real-world Web environments.

Performance declines substantially with task difficulty at the aggregate level. Averaged equally across the 12 models, accuracy drops from 39.66\% on Low to 23.50\% on Medium and 16.76\% on High. GPT-6 Astra declines from 59.26\% to 40.98\% and 24.71\%, while Claude Opus 5.5 achieves 53.70\%, 40.98\%, and 25.88\%, respectively. Model rankings also vary across difficulty levels: GPT-6 Astra leads on Low and ties with Claude Opus 5.5 on Medium, whereas Gemini 3.8 Flash performs best on High at 28.24\%. Qwen3.8-Flash and DeepSeek-V4-Pro achieve only 4.71\% and 3.53\% on High. Although individual models do not always exhibit a monotonic decline across levels, the aggregate decrease indicates that the harder tasks pose a substantial challenge across the evaluated models.

Higher recorded resource consumption does not consistently correspond to higher accuracy. GPT-6 Astra and GPT-5.6 Sol consume nearly identical total numbers of tokens, yet their overall accuracies differ by 11.00 percentage points. Gemini 3.8 Flash consumes approximately 5.22 times as many tokens as GPT-6 Astra but achieves a lower accuracy of 36.50\%. Similarly, DeepSeek-V4-Pro accumulates approximately 2.39 times the total wall-clock time of GPT-6 Astra while achieving only 12.50\% accuracy. These comparisons show that greater time or token consumption alone does not guarantee successful deep search. They motivate further investigation into how models navigate real-world Web sources, locate relevant information, and combine evidence to answer complex questions. Appendix~\ref{sec:more-experiments} reports detailed performance breakdowns by task domain, evidence attributes, and anchor-video duration.


\begin{figure}[!t]
    \centering
    \includegraphics[width=1.0\textwidth]{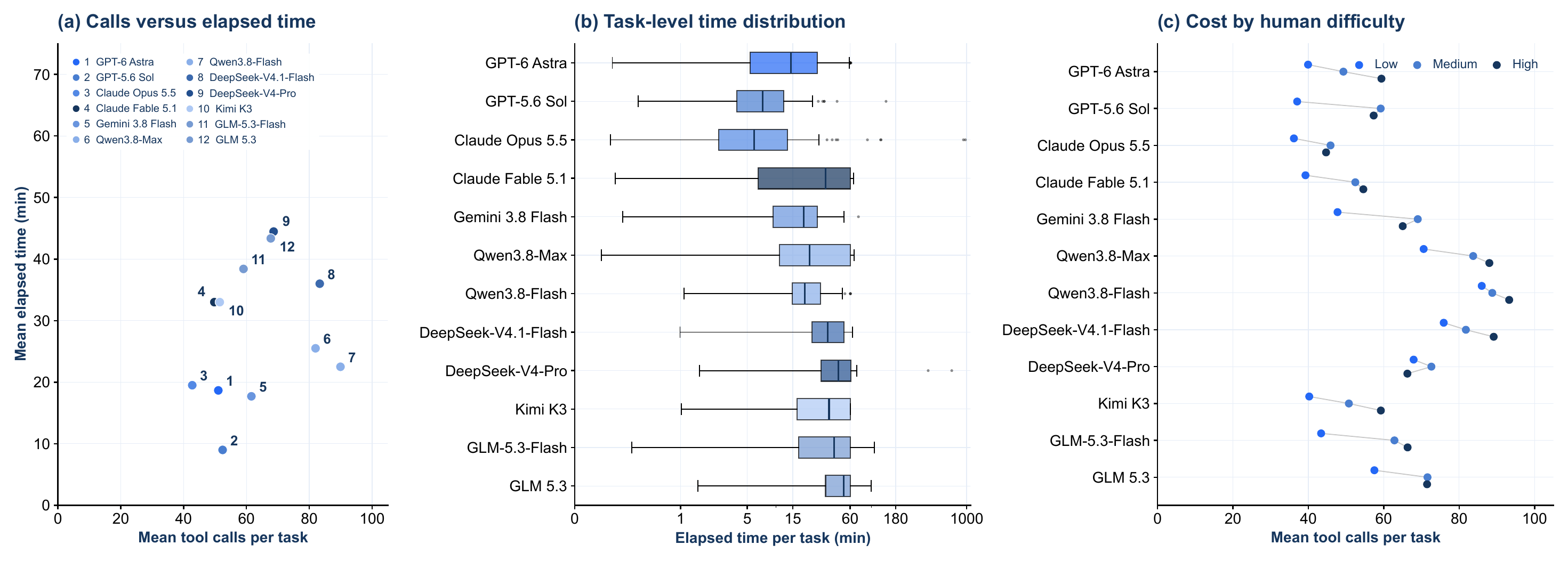}
    \caption{\textbf{Search cost across models and task difficulties.}
    (a) Mean tool calls and elapsed time per target task.
    (b) Distribution of task-level elapsed time, with a logarithmic scale above one minute.
    (c) Mean tool calls per target task within each difficulty level.
    Mean costs are computed from recorded trajectories using the full target-task count as the denominator.}
    \label{fig:cost}
\end{figure}

\subsection{Cost Analysis.} Figure~\ref{fig:cost} shows the search cost of different models in terms of tool calls and wall-clock time. Overall, resource consumption does not exhibit a simple correspondence with final performance: GPT-6 Astra uses 51.05 tool calls and 18.66 minutes per task on average and achieves the highest overall accuracy; GPT-5.6 Sol uses a similar number of calls (52.42) but requires only 8.99 minutes on average. In contrast, Qwen3.8-Flash averages 89.94 tool calls per task, while DeepSeek-V4-Pro and GLM 5.3 require 44.47 and 43.36 minutes, respectively, despite substantially lower accuracy. The task-level runtime distributions show a similar pattern: GPT-5.6 Sol and GPT-6 Astra have median runtimes of 7.22 and 14.35 minutes, respectively, whereas DeepSeek-V4-Pro and GLM 5.3 reach 45.28 and 51.15 minutes, with several models exhibiting pronounced long tails. This suggests that search cost depends on the scale of tool use, as well as whether a model can terminate low-yield trajectories in a timely manner~\citep{wu2025search, qian2026scent}.
Figure~\ref{fig:long-horizon}(a) reports tool-call distributions
by annotated evidence-chain size. Model-level mean call counts are summarized in Figure~\ref{fig:long-horizon}(b). As task difficulty increases, the average number of tool calls rises for all models from Low to Medium; from Medium to High, however, some models continue to increase their tool usage while others plateau or decline. These patterns suggest that tool use on harder tasks is influenced by model-specific search strategies and execution constraints.

\begin{figure}[!t]
    \centering
    \includegraphics[width=\textwidth]{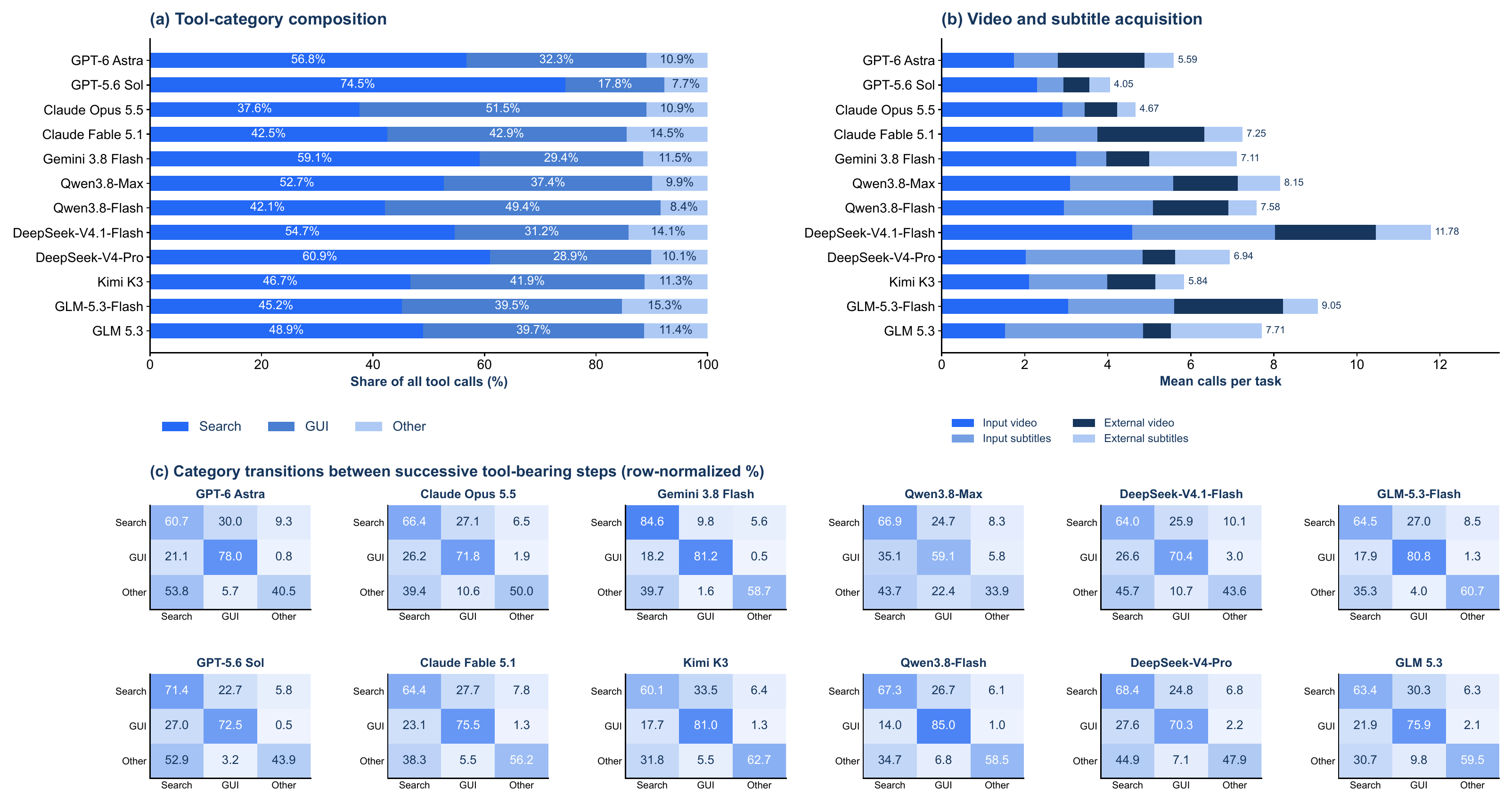}
    \caption{\textbf{Tool-use patterns in real-world Web search.}
    (a) Distribution of recorded tool calls across Search, GUI, and Other categories.
    (b) Mean video and subtitle acquisition calls per target task, separated by input and external sources.
    (c) Row-normalized category transitions between successive tool-bearing steps; rows indicate the current category and columns indicate the next category.}
    \label{fig:tool}
\end{figure}

\begin{figure}[!t]
    \centering
    \includegraphics[width=0.95\textwidth]{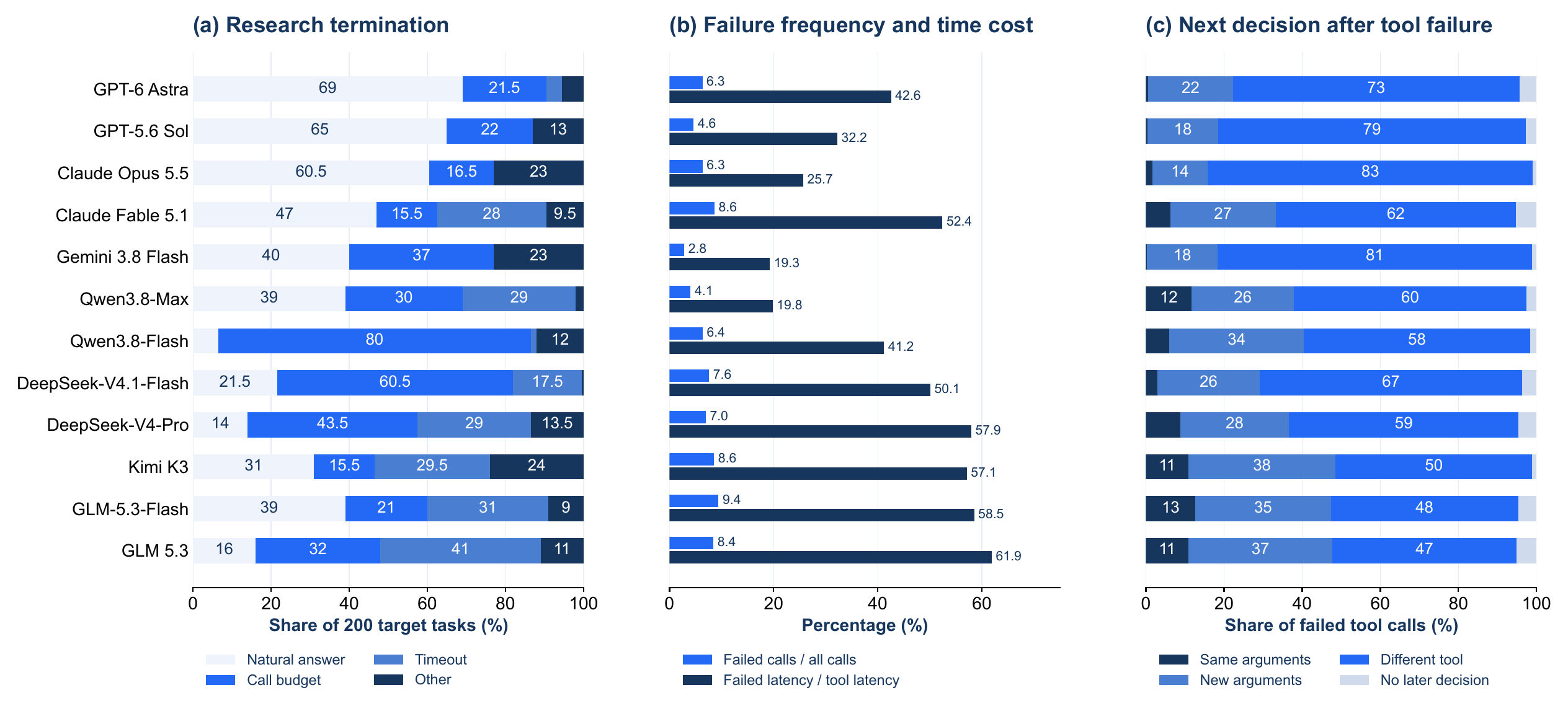}
    \caption{\textbf{Search termination, tool failures, and recovery behavior.}
    (a) Termination outcomes across 200 target tasks. Other includes research terminations caused by incomplete outputs,
    context overflow, or execution errors.
    (b) Failed calls as a percentage of all recorded calls and their share of total recorded tool execution time.
    (c) The next decision following a failed tool call: retrying with the same arguments, modifying arguments, switching tools, or making no subsequent decision.}
    \label{fig:error}
\end{figure}

\subsection{Tool-Use Analysis}

Figure~\ref{fig:tool} shows that models adopt different tool-use patterns under the same tool environment. GPT-5.6 Sol is the most Search-oriented, allocating 74.50\% of its calls to Search and 17.77\% to GUI operations, whereas Claude Opus 5.5 and Qwen3.8-Flash devote 51.52\% and 49.42\% to GUI operations, respectively. However, tool composition alone does not consistently explain performance differences: despite their similar GUI shares, the latter two models achieve overall accuracies of 38.00\% and 12.00\%. More frequent video and subtitle acquisition likewise does not consistently correspond to higher accuracy. DeepSeek-V4.1-Flash averages 11.78 such calls per target task, compared with 4.67 for Claude Opus 5.5, yet achieves lower accuracy (20.50\% versus 38.00\%). Figure~\ref{fig:tool}(c) further shows that within-category transitions generally dominate, with Search$\rightarrow$Search reaching 84.63\% for Gemini 3.8 Flash and GUI$\rightarrow$GUI reaching 84.97\% for Qwen3.8-Flash. These patterns suggest that models often perform consecutive search or interaction steps.

\subsection{Error Analysis}

Figure~\ref{fig:error} summarizes termination outcomes and responses to tool failures. GPT-6 Astra, GPT-5.6 Sol, and Claude Opus 5.5 terminate naturally on 69.0\%, 65.0\%, and 60.5\% of the 200 tasks, respectively. In contrast, Qwen3.8-Flash terminates naturally on only 6.5\%, with 80.0\% reaching the tool-call limit, revealing substantial difficulty in completing search within the execution budget. Meanwhile, \textbf{natural termination and answer correctness remain distinct}: GPT-6 Astra achieves 39.0\% accuracy despite its 69.0\% natural termination rate. Tool failures impose a disproportionate time cost, accounting for only 2.83\%--9.42\% of recorded calls but consuming 19.27\%--61.90\% of recorded tool execution time. For GPT-6 Astra, these proportions are 6.34\% and 42.57\%, respectively. Switching tools is the most common post-failure decision across the evaluated models, accounting for approximately 83\% of cases for Claude Opus 5.5, 81\% for Gemini 3.8 Flash, and 79\% for GPT-5.6 Sol. Kimi K3 and GLM 5.3 devote larger shares to same-tool retries with modified arguments, at approximately 38\% and 37\%. Together, these results identify budget exhaustion, incorrect answers at termination, and costly tool failures as key challenges for deep search in real-world Web environments.

\subsection{Budget Analysis}

\begin{wrapfigure}{r}{0.45\columnwidth}
    \centering
    \includegraphics[width=\linewidth]{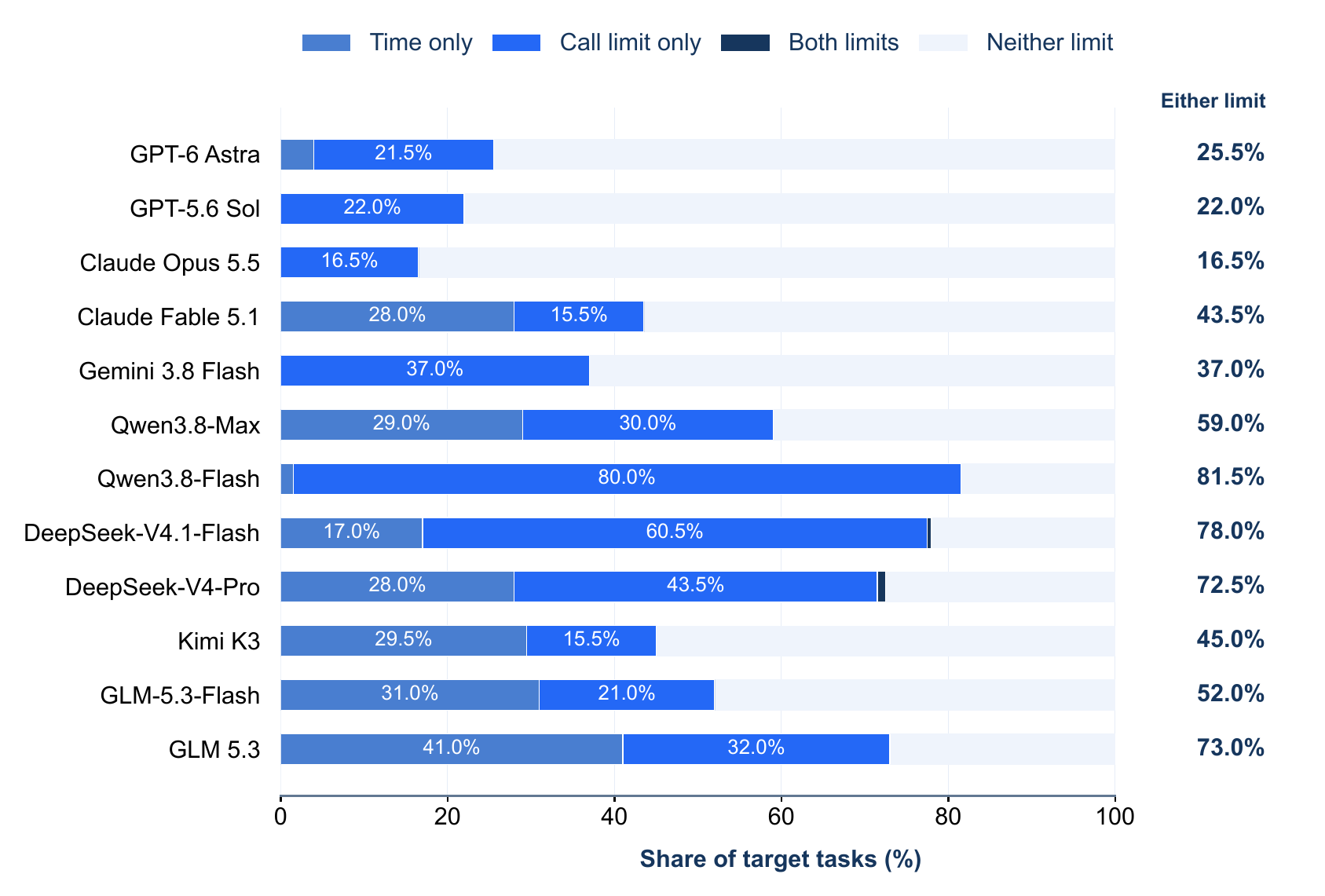}
    \caption{\textbf{Budget exhaustion under the one-hour time limit and the 100-call tool-use limit.}}
    \label{fig:budget-analysis}
\end{wrapfigure}


Figure~\ref{fig:budget-analysis} analyzes budget exhaustion under the one-hour time limit and the 100-call tool-use limit. The proportion of tasks reaching at least one limit varies substantially, from 16.5\% for Claude Opus 5.5, 22.0\% for GPT-5.6 Sol, and 25.5\% for GPT-6 Astra to 81.5\% for Qwen3.8-Flash, 78.0\% for DeepSeek-V4.1-Flash, 73.0\% for GLM 5.3, and 72.5\% for DeepSeek-V4-Pro. Budget exhaustion therefore affects a majority of tasks for several models. The two limits expose different execution patterns. Gemini 3.8 Flash and Qwen3.8-Flash are predominantly call-limited, with call-only rates of 37.0\% and 80.0\%, respectively. DeepSeek-V4.1-Flash similarly reaches the call limit alone on 60.5\% of tasks. In contrast, Claude Fable 5.1, Kimi K3, and GLM 5.3 reach the time limit alone on 28.0\%, 29.5\%, and 41.0\% of tasks, exceeding their respective call-only rates. Simultaneous exhaustion of both budgets occurs on at most 1.0\% of tasks per model, separating time-intensive execution from call-intensive search. Together with Figure~\ref{fig:cost}, these results reveal two central challenges: completing search within the available budget and producing correct answers when search terminates. GPT-6 Astra and Claude Opus 5.5 encounter relatively little budget exhaustion, yet their accuracies remain below 40\%, underscoring the importance of effective source selection, evidence integration, and stopping decisions~\citep{guan2026deeprag, shi2025deepdiver} alongside execution efficiency. Further discussion of long-horizon execution and potential self-improvement directions is provided in Appendix~\ref{sec:discussion}.
\section{Related Work}
\label{sec:related_work}

\paragraph{Deep search in open Web environments.}
Search-augmented models incorporate external retrieval into inference, allowing acquired evidence to guide subsequent reasoning~\citep{li2025search}.
Further work makes these retrieval decisions learnable through reinforcement learning, using outcome feedback to optimize the interaction between search and reasoning~\citep{Jin2025SearchR1TL,chen2025learning}.
In open Web environments, this interaction extends to page navigation and evidence synthesis, with agents learning from Web interaction trajectories~\citep{zheng2025deepresearcher,wu2026webdancer,li2026webthinker}.
Work on adaptive search refines this process by adjusting search frequency and depth to the evidence needed for a query~\citep{shi2025deepdiver}.
As search trajectories lengthen, hierarchical and multi-agent approaches coordinate subgoals and tool execution~\citep{huang2025manusearch,jin2025hira}, while context summarization and state management preserve findings on which later decisions depend~\citep{wu2025resum,qiao2025webresearcher}.
To support training on these longer searches, task-construction methods synthesize increasingly demanding, compositional problems~\citep{tao2026webshaper,lu2025deepdive,liu2025webexplorer}.
Assessing these capabilities requires considering both discovery difficulty and evidence completeness: benchmarks evaluate persistent fact finding and relational complexity~\citep{wei2025browsecomp,zhao2026lohosearch}, and extend assessment to the depth and breadth of evidence collection~\citep{gupta2026deepsearchqa,lan2025deepwidesearch}.
Representing findings as claims makes the information established by a search explicit~\citep{java2026characterizing}, while controlled corpora help examine retrieval and reasoning under shared evidence conditions~\citep{chen2025browsecomp,hu2026sage}.
When the intended outcome is a synthesis, report-level criteria also assess its quality, evidence coverage, and citation support~\citep{du2026deepresearch,li2026deepresearch}.

\paragraph{Grounding deep search in multimodal evidence.}
Multimodal deep search extends evidence gathering to visual clues that must be connected with external Web information.
To support this connection, multimodal search methods learn how to coordinate visual reasoning with image and text retrieval~\citep{geng2026webwatcher,wu2025mmsearch}.
Corresponding evaluation makes the role of perception explicit through required evidence from Web images and videos and verified checklists of multimodal dependencies~\citep{li2025mm}.
Video-grounded methods and benchmarks extend this connection to clues distributed across frames and open-Web retrieval~\citep{liu2026watching,fang2026video}, while work grounded in egocentric video carries visual understanding into subsequent online planning and action~\citep{yu2026ego2web}.
In these settings, interpreting a visual observation can change what an agent searches for next and how it uses the resulting evidence.

\paragraph{Acquiring evidence through interactive access.}
Following these clues on real websites also requires deciding how to access the underlying evidence.
Work on website traversal extends retrieval beyond landing pages by navigating site structure~\citep{wu2025webwalker}, while unindexed information seeking broadens these access requirements to dynamic interfaces and embedded documents~\citep{liu2026uis}.
To execute such operations reliably, browser-agent designs strengthen element grounding and task-state management~\citep{li2026avenir,chen2026wuying}.
When access spans multiple pages or websites, validated task generation~\citep{huang2026gta} and compositional cross-site evaluation~\citep{tian2025mmina,xu2026cap} examine the coordination of navigation, perception, and dependent subtasks.
Assessing whether these interactions complete the task also draws on intermediate visual evidence and action histories from live websites~\citep{xue2025illusion}.
These efforts make access operations part of deep search, since a relevant source may require additional navigation, interface manipulation, or artifact processing before its evidence becomes usable.
\par

\section{Conclusion}

We introduce OpenDS-World, a benchmark for deep search in real-world Web environments. Its 200 tasks originate from audiovisual anchors and require discovering and connecting evidence across heterogeneous online resources. The annotated evidence chains cover four modalities, 42 formats, and seven access types, capturing the diversity of both Web information and the interactions needed to obtain it. Evaluations of 12 frontier models reveal substantial challenges: the strongest model achieves only 39.00\% accuracy, and compact evidence chains expand into lengthy execution trajectories as agents resolve uncertain clues and explore candidate sources. Further analyses reveal costly tool failures, frequent budget exhaustion for several models, and a persistent gap between natural termination and answer correctness. OpenDS-World provides a concrete foundation for studying long-horizon search and recursive self-improvement, with the goal of developing agents that can reliably discover, access, and integrate evidence in real-world Web environments.


\makeatletter

\renewcommand\@biblabel[1]{[#1]}
\let\@bibsetup\NAT@bibsetnum

\makeatother

\newpage

\bibliographystyle{plainnat}
\bibliography{reference}

\newpage

\newpage

\appendix
\crefalias{section}{appendix}
\crefname{appendix}{Appendix}{Appendices}

\addcontentsline{toc}{section}{Appendix}
\begin{center}
    {\Large\bfseries Appendix}
\end{center}


\section{Acknowledgments}
\label{sec:acknowledgments}

We thank all participants in the human evaluation of OpenDS-World for their time and effort in solving the tasks and assessing their difficulty. Their contributions supported the human performance baseline and our analysis of the challenges posed by real-world deep search. We gratefully acknowledge the following participants, listed in no particular order:

Luoshi Liang, Lida Zhang, Ruining Ma, Qianrui Cui,
Haihan Liu, Jinbo Chen, Lin Shi, Ximing Liu,
Siyuan Yu, Siyuan Feng, Yanjie Huang, Xintong Tan,
Yingxin Fu, Hao Ouyang, Zhenye Su, and Qile Li; Leyi Sheng, Yue Guo, Guangyu Wei, Yangyang Li, Gutai Fu, Jiajun Yuan, Dongming Han, Cheng Peng, Xinyue Tang, Fenglei Zhou, Chuanluan Xu, Chonggao Yin, Zidong Nie, Hanqi Zhang, and Xiang Li; Yuzhe Zhuang, Xixi Miao, Ruilang Chen, Haoran Wang, Lehao Chen, Jiarui Liang, Changhao Liu, Gang Feng, Xiaosi Xu, Dongsheng Yang, Hongliang Chen, Qixiao Sun,
Chenhui Zhang, Yingqi Hou, and an anonymous participant. 

Participants are affiliated with Hong Kong University of Science and Technology (Guangzhou), South China Normal University, Ocean University of China, University of South China, Qingdao University of Science and Technology, and Henan University.

\section{OpenDS-World Construction Details}
\label{sec:opends-construction}

\subsection{Benchmark Composition}
\label{sec:dataset-composition}

\begin{figure}[tbp]
    \centering
    \includegraphics[width=\linewidth]{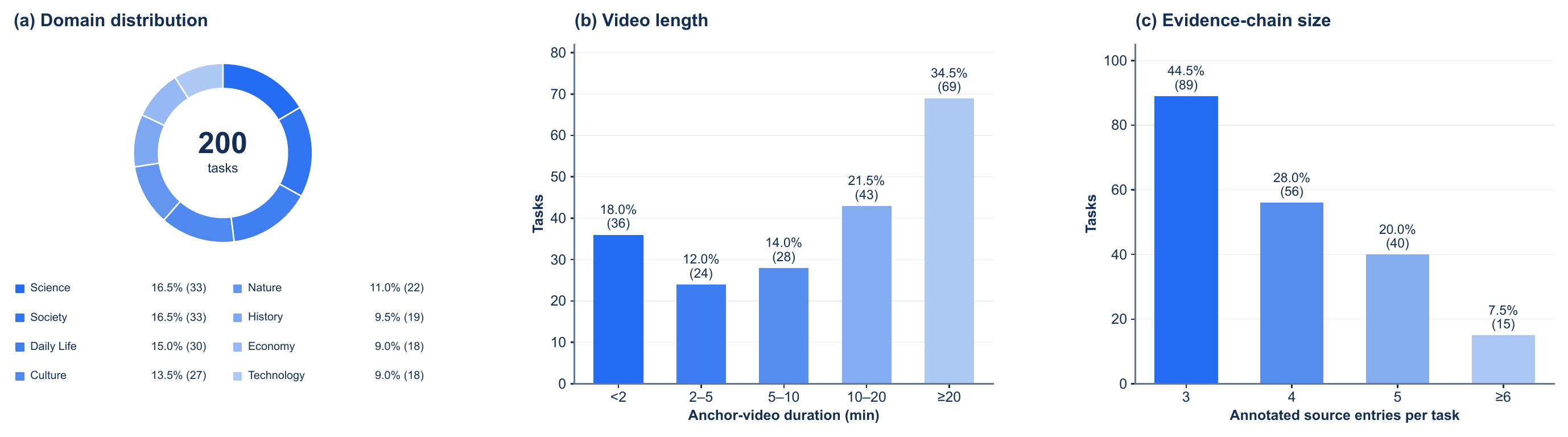}
    \caption{\textbf{Task composition and annotated evidence structure of OpenDS-World.}
    (a) Distribution of 200 tasks across eight domains.
    (b) Distribution of anchor-video durations across five temporal intervals.
    (c) Number of annotated source entries per task, with six or more entries grouped into the final bin.
    Percentages are calculated over all 200 tasks, with task counts reported in parentheses.
    Evidence-chain statistics exclude the separately supplied anchor video.}
    \label{fig:app-S01_dataset}
\end{figure}

Figure~\ref{fig:app-S01_dataset} presents the task composition and annotated evidence structure of OpenDS-World, covering domain distribution, anchor-video duration, and evidence-chain size. As shown in Figure~\ref{fig:app-S01_dataset}(a), the benchmark comprises 200 tasks spanning eight domains, each accounting for 9.0\% to 16.5\% of the dataset. Science and Society contribute 33 tasks each, followed by Daily Life (30), Culture (27), Nature (22), History (19), Economy (18), and Technology (18). This distribution provides broad coverage across specialized knowledge and everyday information needs, without any single domain dominating the benchmark.

Figure~\ref{fig:app-S01_dataset}(b) illustrates the distribution of anchor-video durations. The videos range from short clips to extended recordings, with 18.0\% of tasks featuring videos shorter than two minutes, 12.0\% between two and five minutes, 14.0\% between five and ten minutes, 21.5\% between ten and twenty minutes, and 34.5\% lasting at least twenty minutes. Notably, 56.0\% of tasks begin with videos of at least ten minutes, requiring agents to identify relevant clues within potentially extensive temporal content.

Figure~\ref{fig:app-S01_dataset}(c) shows the distribution of annotated evidence-chain sizes. The 200 tasks contain a total of 787 verified source entries, averaging 3.94 entries per task. Specifically, 44.5\% of tasks contain three source entries, 28.0\% contain four, 20.0\% contain five, and 7.5\% contain six or more. Thus, 55.5\% of tasks require at least four annotated source entries. These entries constitute the verified evidence supporting the reference answers, while agents must independently discover the relevant sources and establish their connections during execution.   

\subsection{Evidence Attribute Diversity}
\label{sec:evidence-diversity}

\begin{figure}[tbp]
    \centering
    \includegraphics[width=\linewidth]{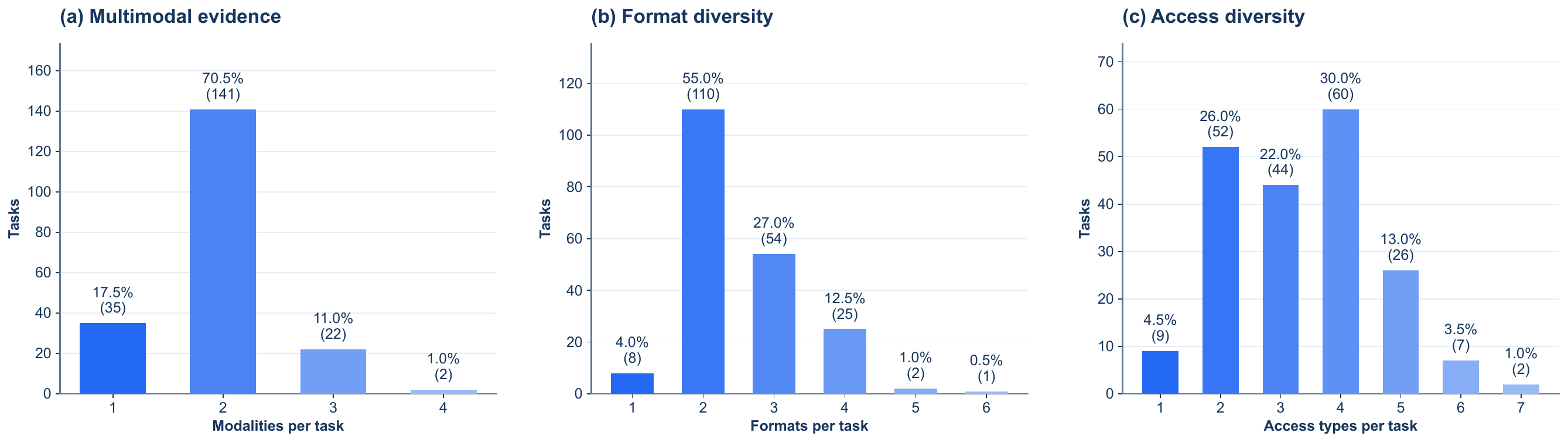}
    \caption{\textbf{Evidence attribute diversity within individual tasks.}
    (a) Number of distinct information modalities per task.
    (b) Number of distinct resource formats per task.
    (c) Number of distinct access requirement types per task.
    Each attribute is counted at most once within a task.
    Percentages are calculated over all 200 tasks, with task counts reported in parentheses.
    Statistics exclude the separately supplied anchor video.}
    \label{fig:app-S02_diversity}
\end{figure}

Beyond the overall evidence-chain size, OpenDS-World exhibits substantial diversity in information modalities, resource formats, and access requirements. Figure~\ref{fig:app-S02_diversity} characterizes this diversity by counting the number of distinct attributes within each task. Each attribute is counted at most once per task, regardless of how many annotated source entries exhibit it. All statistics are based on annotated evidence sources and exclude the separately supplied anchor video.

\paragraph{Multimodal Evidence.}
As shown in Figure~\ref{fig:app-S02_diversity}(a), 17.5\% of tasks involve a single information modality, while 70.5\% involve two modalities, 11.0\% involve three, and 1.0\% involve all four modalities. Consequently, 82.5\% of tasks require evidence spanning at least two modalities, with bimodal evidence being the most common configuration. This distribution highlights the prevalence of multimodal evidence integration, where agents must establish connections across information presented in textual, visual, audio, and video forms.

\paragraph{Format Diversity.}
Figure~\ref{fig:app-S02_diversity}(b) presents the number of distinct resource formats involved in each task. Only 4.0\% of tasks involve a single format, whereas 55.0\% involve two formats, 27.0\% involve three, and 12.5\% involve four. The remaining 1.5\% involve five or six formats. Overall, 96.0\% of tasks require at least two resource formats, and 41.0\% involve three or more. Unlike modality diversity, format diversity reflects differences in resource representation and processing requirements. For instance, HTML, PDF, and CSV may all contain textual information but require different access and parsing mechanisms. The observed distribution therefore emphasizes the heterogeneity of resource handling in real-world Web research.

\paragraph{Access Diversity.}
Figure~\ref{fig:app-S02_diversity}(c) illustrates the number of distinct access requirement types associated with each task. The most common configuration involves four access types, accounting for 30.0\% of tasks, followed by two types (26.0\%), three types (22.0\%), and five types (13.0\%). Overall, 95.5\% of tasks involve at least two access types, 69.5\% involve at least three, and 47.5\% involve four or more. Two tasks require all seven access types. These results demonstrate that evidence acquisition frequently involves combinations of heterogeneous operations. 

\subsection{Long-Tail File Formats}
\label{sec:long-tail-formats}

In addition to the seven most frequent resource formats reported in the main paper, OpenDS-World contains 108 source entries spanning less common formats, accounting for 13.72\% of all 787 annotated entries. Table~\ref{tab:app-long-tail-formats} groups these long-tail formats by resource category. They encompass structured Web data, spreadsheets, scientific datasets, documents, geospatial data, compressed files, and 3D assets. These resources differ in their data structures and processing requirements, and accessing their evidence may involve Web navigation, resource downloading, and specialized file parsing.

For consistency, Table~\ref{tab:app-long-tail-formats} standardizes resource format names while preserving the original annotation counts. C++ source files, phylogenetic trees, and RDF data are grouped with plain-text resources under Other text files.

\begin{table}[tbp]
\centering
\caption{\textbf{Long-tail resource formats and file types.}
Resources categorized as Other formats in the main paper are
grouped by resource family.
Numbers in parentheses indicate entry counts, and percentages
are calculated over all 787 annotated source entries.
The Other text files category retains the original annotations.}
\label{tab:app-long-tail-formats}

\small
\setlength{\tabcolsep}{5pt}
\renewcommand{\arraystretch}{1.2}

\begin{tabular}{@{}p{0.17\linewidth}p{0.55\linewidth}rr@{}}
\toprule
\textbf{Resource group} &
\textbf{Formats} &
\textbf{Entries} &
\textbf{Share} \\
\midrule
Web &
JSON (11), XHTML (5), XML (5), RSS (4), JSONL (1) &
26 & 3.30\% \\
Images &
WebP (9), PNG (4), TIFF (4), EPS (2) &
19 & 2.41\% \\
Tables &
XLSX (7), Parquet (4), TSV (3), ODS (2) &
16 & 2.03\% \\
Scientific data &
FASTA (5), DICOM (3), ROOT (3), SDF (3), CIF (1) &
15 & 1.91\% \\
Documents &
PPTX (3), R Markdown (3), RTF (3), EPUB (2),
LaTeX (2), Markdown (1) &
14 & 1.78\% \\
Other text files &
Plain text (2), C++ source (2),
phylogenetic trees (3), RDF data (1) &
8 & 1.02\% \\
Geospatial data &
KML (3), GeoJSON (1) &
4 & 0.51\% \\
Compressed files &
TAR.GZ (2), GZIP (1) &
3 & 0.38\% \\
3D assets &
GLB (1), OBJ (1) &
2 & 0.25\% \\
Audio &
M4A (1) &
1 & 0.13\% \\
\midrule
\textbf{Total} & & \textbf{108} & \textbf{13.72\%} \\
\bottomrule
\end{tabular}
\end{table}

\subsection{Access Requirement Taxonomy}
\label{sec:access-taxonomy}

To characterize the diversity of evidence access mechanisms in real-world Web environments, we classify access requirements into seven types according to the operations needed to obtain the target evidence, as summarized in Table~\ref{tab:access-taxonomy}. These types cover immediate observation, hypermedia transitions, parameterized state updates, temporal indexing, graphical state manipulation, declarative queries, and artifact transfer. Together, they describe the principal operations involved in accessing evidence across heterogeneous Web sources.

\begin{table}[tbp]
\centering
\caption{\textbf{Taxonomy of access requirements.}
The seven access types characterize distinct operations
involved in obtaining annotated evidence.
Multiple labels may apply to the same source entry.}
\label{tab:access-taxonomy}

\small
\setlength{\tabcolsep}{4pt}
\renewcommand{\arraystretch}{1.2}

\begin{tabular}{@{}lp{0.23\linewidth}p{0.65\linewidth}@{}}
\toprule
\textbf{Code} &
\textbf{Access type} &
\textbf{Definition} \\
\midrule
O &
Immediate Observation &
Observe the target evidence directly in the current state,
without changing the source state. \\
N &
Hypermedia Transition &
Change location through discrete navigation within
or across pages and documents. \\
P &
Parameterized State Update &
Change the source state by entering or selecting
explicit parameters. \\
T &
Temporal Indexing &
Locate content along the temporal dimension
of a source. \\
G &
Graphical State Manipulation &
Expose information through interaction with
a graphical or spatial representation. \\
Q &
Declarative Query &
Express information needs through a structured
query language or machine interface. \\
X &
Artifact Transfer &
Transfer an information-bearing artifact from a remote
environment into an environment accessible for
agent processing. \\
\bottomrule
\end{tabular}
\end{table}

The access labels are not mutually exclusive and can be combined to describe multi-operation evidence acquisition. For example, retrieving evidence from a downloadable report may involve following a link (N), transferring the file (X), and inspecting its contents (O). Similarly, locating and examining a specific video segment may require both temporal indexing (T) and immediate observation (O). For task-level statistics, we take the union of the access labels assigned to all annotated source entries within each task, counting each type at most once per task. Consequently, coverage percentages across access types may sum to more than 100\%.

Importantly, this taxonomy is defined at the level of access operations, independent of the specific tools used to perform them. For instance, parameterized state updates modify explicit input values or selections, whereas graphical state manipulation changes the spatial or visual representation through interaction. Declarative queries specify target records or retrieval conditions through structured interfaces. A single tool may support multiple access types, while a particular access operation may require multiple tool calls.

These annotations describe task-side access requirements along the verified evidence paths and form the basis for the coverage statistics and difficulty-conditioned analysis presented below. Agent actions are derived separately from execution trajectories.

\subsection{Source Requirements Across Difficulty Groups}
\label{sec:source-requirements-difficulty}

\begin{figure}[tbp]
    \centering
    \includegraphics[width=\linewidth]{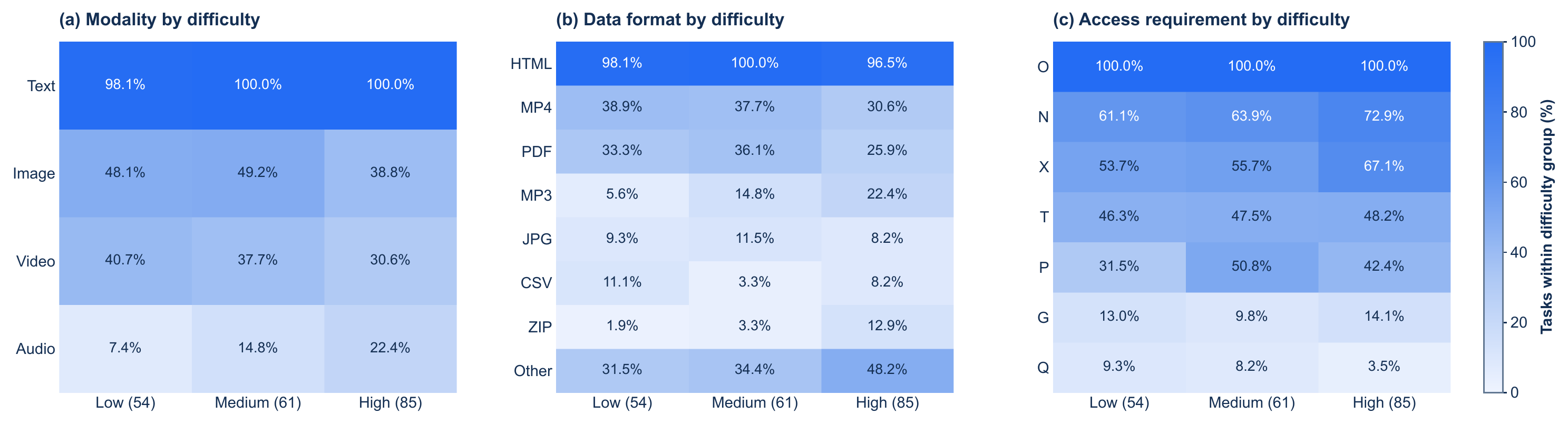}
    \caption{\textbf{Source requirements across difficulty groups.}
    (a) Information modality coverage.
    (b) Resource format coverage.
    (c) Access requirement coverage.
    Percentages are calculated within the human-calibrated
    Low (54 tasks), Medium (61 tasks), and High (85 tasks)
    groups.
    A task may involve multiple modalities, resource formats,
    and access types.
    Statistics are based on annotated source entries and
    exclude the separately supplied anchor video.}
    \label{fig:app-S03_attributes_by_difficulty}
\end{figure}

Figure~\ref{fig:app-S03_attributes_by_difficulty} compares the distributions of information modalities, resource formats, and access requirements across human-calibrated difficulty groups. Based on the independent human evaluation, the 200 tasks are divided into three groups: Low comprises 54 tasks solved by at least two human participants, Medium comprises 61 tasks solved by exactly one participant, and High comprises 85 tasks that none of the five participants solved. These groups are defined by observed human solving success as the sole difficulty criterion. Coverage percentages are calculated separately within each group to account for differences in group size.

The distributions reveal several differences in evidence composition across difficulty groups. As shown in Figure~\ref{fig:app-S03_attributes_by_difficulty}(a), audio coverage increases from 7.4\% in Low to 22.4\% in High, whereas image coverage decreases from 48.1\% to 38.8\% and video coverage decreases from 40.7\% to 30.6\%. The resource format distributions in Figure~\ref{fig:app-S03_attributes_by_difficulty}(b) also vary across groups. For example, ZIP coverage increases from 1.9\% in Low to 12.9\% in High, while the coverage of formats grouped as Other increases from 31.5\% to 48.2\%.

Differences are also observed in access requirements. As illustrated in Figure~\ref{fig:app-S03_attributes_by_difficulty}(c), hypermedia transition (N) coverage increases from 61.1\% in Low to 72.9\% in High, while artifact transfer (X) coverage rises from 53.7\% to 67.1\%. In contrast, temporal indexing (T) remains relatively stable at approximately 47\%--48\% across all three groups. Other access types exhibit non-monotonic patterns, indicating that their prevalence does not consistently increase with task difficulty.

Overall, the results show that difficulty groups differ in their combinations of evidence modalities, resource formats, and access requirements. Compared with the Low group, the High group shows higher coverage of audio, ZIP archives, hypermedia transitions, and artifact transfers, and lower coverage of image and video evidence.

\section{Detail Experiment Setup}
\label{sec:execution-environment}

\begin{longtable}{p{.31\textwidth}p{.29\textwidth}p{.32\textwidth}}
\caption{Tool interfaces. Optional arguments and constraints are defined by the implementation.}
\label{tab:tool}\\
\toprule
Tool & Arguments & Function \\ \midrule \endfirsthead
\toprule Tool & Arguments & Function \\ \midrule \endhead

\multicolumn{3}{@{}l}{\textbf{Search}}\\

\texttt{text\_\allowbreak{}search} & query & Search the Web for textual results. \\
\texttt{image\_\allowbreak{}search} & query & Search for images relevant to a query. \\
\texttt{open\_\allowbreak{}page} & url & Retrieve the contents of a Web page. \\
\texttt{youtube\_\allowbreak{}search} & query; max\_\allowbreak{}results & Search videos; return at most ten results. \\

\multicolumn{3}{@{}l}{\textbf{GUI}}\\

\texttt{browser\_\allowbreak{}navigate} & url & Navigate the interactive browser. \\
\texttt{browser\_\allowbreak{}get\_\allowbreak{}markdown} & none & Read the current page as Markdown. \\
\texttt{browser\_\allowbreak{}get\_\allowbreak{}clickable\_\allowbreak{}elements} & none & List indexed interactive elements. \\
\texttt{browser\_\allowbreak{}click} & index & Click an indexed element. \\
\texttt{browser\_\allowbreak{}form\_\allowbreak{}input\_\allowbreak{}fill} & index; value; clear & Fill a browser input element. \\
\texttt{browser\_\allowbreak{}press\_\allowbreak{}key} & key & Send a keyboard action. \\
\texttt{browser\_\allowbreak{}scroll} & amount & Scroll the page by a signed amount. \\
\texttt{browser\_\allowbreak{}screenshot} & name & Capture the current browser view. \\
\texttt{browser\_\allowbreak{}get\_\allowbreak{}download\_\allowbreak{}list} & path & List downloaded artifacts. \\
\texttt{browser\_\allowbreak{}get\_\allowbreak{}download\_\allowbreak{}content} & path; query; line/column coordinates & Read or search a downloaded artifact. \\

\multicolumn{3}{@{}l}{\textbf{Other}}\\

\texttt{youtube\_\allowbreak{}read\_\allowbreak{}video} & url\_\allowbreak{}or\_\allowbreak{}id; num\_\allowbreak{}frames; start\_\allowbreak{}seconds; end\_\allowbreak{}seconds & Inspect sampled video frames globally or within a time window. \\
\texttt{youtube\_\allowbreak{}read\_\allowbreak{}subtitle} & url\_\allowbreak{}or\_\allowbreak{}id; start\_\allowbreak{}seconds; end\_\allowbreak{}seconds & Read subtitle content, optionally restricted to a time window. \\
\bottomrule\end{longtable}

\subsection{Setting}

\subsubsection{Tool Setting} \label{sec:tool-interfaces}

Table~\ref{tab:tool} summarizes the 16 tools available in the OpenDS-World evaluation environment, including their input parameters and supported operations. Browser interactions are conducted within a persistent session, where agents identify interaction targets using element indices returned from the current page state. For non-Web resources, the environment supports access to downloaded artifacts through task-local file paths, as well as video frame sampling and subtitle extraction within specified temporal intervals.

\subsubsection{Input Videos Process}

To ensure consistency and reproducibility of task inputs, OpenDS-World preserves the initial videos as read-only files and makes them accessible to models through the predefined identifier \texttt{input\_video}, avoiding dependence on the availability of the original YouTube videos during evaluation. Models can inspect either the entire video or specific temporal intervals through a unified video-reading interface, while external videos are accessed via URLs or video identifiers. The video reader samples 10 frames by default and supports 1 to 100 frames per call, sampling at the midpoints of equally spaced temporal intervals. The same sampling strategy applies to inspection within a specified time range. Additionally, a separate subtitle-reading interface supports the extraction of timestamped subtitles from the entire video or a specified temporal interval.

\subsection{Agent Execution}
\label{sec:agent-execution}

\subsubsection{Budget Management}
\label{sec:budget-management}

Table~\ref{tab:agent-execution-settings} summarizes the execution budgets and key configuration parameters of the OpenDS-World evaluation harness. Research budgets follow the unified evaluation protocol described in the main text, while the remaining parameters reflect the current implementation defaults. Visual observations are managed separately. When the number of images exceeds the configured limit in Table~\ref{tab:agent-execution-settings}, older images are progressively replaced with textual omission markers, while the original images remain preserved in the execution trajectories for subsequent analysis. Historical experiments used image limits of 250, with the actual setting determined by the corresponding run configuration.

\begin{table}[t]
\centering
\caption{Agent execution and context management settings.}
\label{tab:agent-execution-settings}
\small
\setlength{\tabcolsep}{5pt}
\renewcommand{\arraystretch}{1.15}
\begin{tabular}{@{}p{0.43\linewidth}p{0.49\linewidth}@{}}
\toprule
Setting & Value or Behavior \\
\midrule
Research time budget & 3,600 seconds \\
Research tool-call budget & 100 calls \\
Recent tool observations & Latest 4 exempt from text compaction \\
Older search observations & Retain the first 2 results \\
Older textual observations & Retain up to 2,000 characters \\
Image-context limit & 250 images by default \\
Finalization output budget & 5,000 tokens across at most 2 requests \\
\bottomrule
\end{tabular}
\end{table}

\subsubsection{Context Management}
\label{sec:context-management}

To control context growth during long search trajectories, the harness applies deterministic compaction to older observations before each model request, without relying on an additional summarization model. As shown in Table~\ref{tab:agent-execution-settings}, the four most recent tool observations remain intact. For older observations, search-result fields retain only the first two entries, while textual content fields are truncated to 2,000 characters. Other fields remain unaffected by these limits.

\subsubsection{Termination Mechanism}
\label{sec:termination-mechanism}

\noindent\textbf{Research Termination.}
The research phase terminates naturally when the model
produces a complete, nonempty answer without requesting
additional tool calls. It may also terminate due to budget
exhaustion, incomplete output, context overflow, or
execution errors. The harness records the research
termination reason and answer-generation status separately,
while answer correctness is determined by the independent
evaluation procedure.

\noindent\textbf{Final Answer Generation.}
If the research phase ends without a complete answer,
the harness enters a separate finalization stage using
the evidence collected during research. No tools are
available during this stage, preventing further searches
or access to external resources.
As shown in Table~\ref{tab:agent-execution-settings},
the default configuration permits at most two model
requests sharing a 5,000-token output budget, with
4,000 tokens allocated to the first request.
Only nonempty, complete responses without tool calls
are accepted, and the finalization outcome is recorded.

\subsection{Prompts}
\label{app:agent-prompts}

\subsubsection{Search Prompts}
\label{sec:prompts-evaluation}

OpenDS-World employs a unified system prompt to guide agents in conducting autonomous Web research, inspecting input media, and verifying supporting evidence. The prompt does not prescribe search strategies, source selection, or tool-call sequences, leaving these decisions to the models. The complete system prompt is provided below.

\begin{promptbox}{Research System Prompt}
You are participating in a controlled live-web research benchmark.

Use only the supplied tools for web, browser, download, and YouTube access. Conduct thorough, autonomous research and use the available tools as you judge appropriate. Inspect input media when present, verify the information needed to answer the question, and do not fabricate unsupported facts. Decide your own research strategy and continue until you have sufficient evidence or the available budget is exhausted.

Final-answer format: follow any explicit format requested in the question. Otherwise, output only one short English paragraph, normally one to three sentences and no more than 80 words. State the answer directly. Do not include headings, bullet points, a research narrative, chain-of-thought, or labels such as "Final Answer".
\end{promptbox}

Each task provides the agent with a question and an input video accessible through the reserved identifier \texttt{input\_video}, without annotated intermediate evidence or the reference answer. The task-input template is provided below.

\begin{promptbox}{Task Input Template}
Task-provided local video access ID: input_video
Recommended: first use youtube_read_video with url_or_id="input_video" to inspect this complete local video file. This reserved ID is not a YouTube search-result ID and does not require network access. Start with a global overview by omitting start_seconds and end_seconds. You may also use youtube_read_subtitle with the same reserved ID to read its subtitle track independently.

Question:
[Task question]
\end{promptbox}

\subsubsection{Final Answer Prompt}

When the research phase ends without a complete answer, the harness employs a separate final-answer prompt, instructing the model to respond solely based on the evidence already collected, without further tool calls. The complete prompt is provided below.

\begin{promptbox}{Final Answer Prompt}
Research has ended. Using only the completed evidence in this conversation, give the best-supported final answer now. Do not call tools or invent missing facts. If the evidence is insufficient, say so briefly. Follow any answer format specified in the question; otherwise give one short English paragraph of 1-3 sentences, at most 80 words. Do not include a research narrative or headings. Keep deliberation brief enough to leave room for the final answer.
\end{promptbox}

\subsection{LLM as Judge Evaluation}
\label{app:judge-evaluationllm}

\paragraph{Reference-based Judgment.}
OpenDS-World uses Qwen3.8-27B~\citep{qwen38} as the answer judge, evaluating candidate answers against reference answers based on semantic equivalence and completeness~\citep{wei2025browsecomp, liu2026watching}. The judge relies on the reference answer without access
to external tools or independently solving the question. The complete judge system prompt is provided below.

\begin{promptbox}{Answer Judge System Prompt}
You are a strict benchmark answer judge. Compare the candidate answer with the authoritative reference answer, using the question only to understand what must be answered.

Rules:
- Mark correct only when the candidate is semantically equivalent to the reference and answers every required part.
- Numbers, named entities, dates, directions, and list membership must be correct.
- Extra explanation is allowed only if it does not contradict the reference.
- A refusal, an insufficient-evidence response, an empty answer, or an answer from a failed execution is incorrect.
- Do not solve the question independently and do not use outside knowledge. Treat the reference answer as authoritative.

Return exactly one JSON object with this schema:
{"correct": true, "confidence": 0.0, "reason": "brief comparison"}
\end{promptbox}

The judge receives the question, reference answer, candidate answer, and execution status in the following JSON format.

\begin{promptbox}{Answer Judge Input Format}
{
  "question": "[Task question]",
  "reference_answer": "[Reference answer]",
  "candidate_answer": "[Candidate answer]",
  "execution_status": "[Recorded execution status]",
  "execution_error": null
}
\end{promptbox}

An absent candidate answer is represented as \texttt{(empty)}, while \texttt{execution\_error} is set to \texttt{null} when no error is recorded.


\section{More Experiments}
\label{sec:more-experiments}

This section presents a detailed analysis of model performance across task domains, evidence attributes, and video duration.

\subsection{Domain Performance Analysis}
\label{sec:domain-analysis}

\begin{figure}[tbp]
\centering
\includegraphics[width=0.85\linewidth]{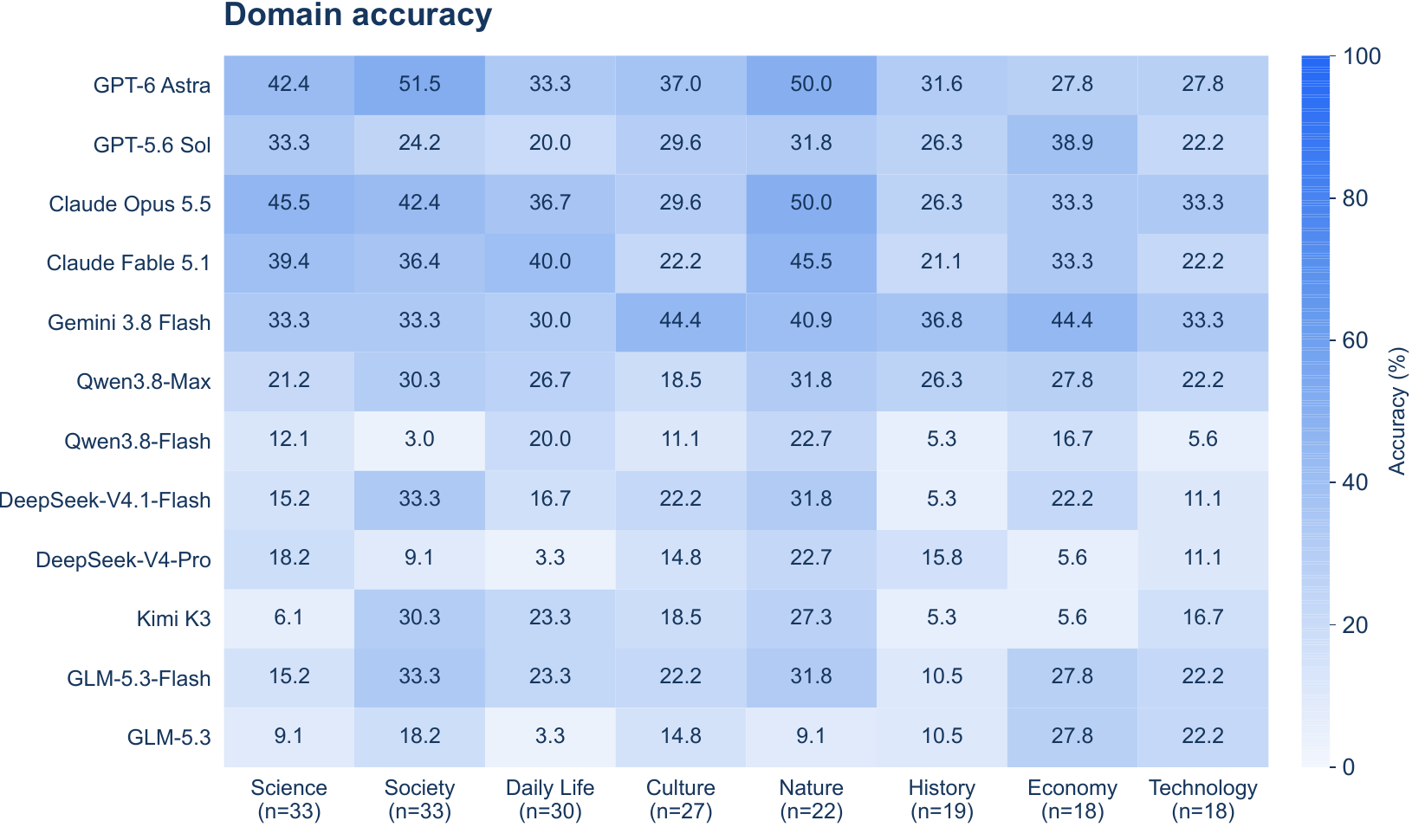}
\caption{Domain breakdown for all 12 models. Columns are mutually exclusive task domains; values denote accuracy (\%).}
\label{fig:fit-domain}
\end{figure}

Figure~\ref{fig:fit-domain} presents the accuracy distribution of 12 models across eight task domains. The models exhibit distinct domain-specific performance patterns. GPT-6 Astra achieves relatively high accuracy in Society (51.5\%) and Nature (50.0\%), while Claude Opus 5.5 performs particularly well in Nature (50.0\%) and Science (45.5\%). In contrast, Gemini 3.8 Flash achieves 44.4\% accuracy in both Culture and Economy, and attains the highest accuracy in History (36.8\%). Across domains, Nature yields relatively high accuracy for most models. Cross-model variation is particularly pronounced in Science and Society, which have the largest standard deviations in accuracy (13.2 and 12.9 percentage points, respectively). Overall, the relative strengths of different models vary across task domains, revealing heterogeneous domain-level performance patterns.

\subsection{Evidence Attributes Performance Analysis}

\paragraph{Modality.}
Table~\ref{tab:fit-modality} presents model accuracy across four evidence modalities: Text (199 tasks), Image (89 tasks), Video (71 tasks), and Audio (32 tasks). GPT-6 Astra achieves the highest accuracy on Text (38.69\%) and Image (48.31\%), while Gemini 3.8 Flash leads on Video (43.66\%) and shares the highest Audio accuracy (50.00\%) with Claude Fable 5.1. Claude Opus 5.5 also demonstrates strong performance on Image (44.94\%) and Audio (46.88\%). Notably, Gemini 3.8 Flash performs better on Video (43.66\%) and Audio (50.00\%) than on Text (36.68\%), whereas GPT-6 Astra achieves its highest accuracy on Image (48.31\%). Overall, the results reveal distinct performance patterns across evidence modalities, with different models exhibiting relative strengths in different modality groups.

\begin{table}[t]
\centering
\caption{\textbf{Modality results on OpenDS-World.}
We report accuracy (\textbf{Acc}, \%) across tasks involving text, image, video, and audio evidence.
Numbers in parentheses indicate correct answers;
bold accuracy values denote the best result in each modality.
A task may belong to multiple modality categories.}
\label{tab:fit-modality}

\begingroup
\small
\renewcommand{\arraystretch}{1.26}
\setlength{\tabcolsep}{7pt}
\setlength{\aboverulesep}{3pt}
\setlength{\belowrulesep}{3pt}
\setlength{\heavyrulewidth}{0.8pt}
\setlength{\lightrulewidth}{0.4pt}

\resizebox{0.6\textwidth}{!}{%
\begin{tabular}{
@{}l
r
@{\hspace{12pt}}r
@{\hspace{12pt}}r
@{\hspace{12pt}}r
@{}
}
\toprule

& \multicolumn{1}{c}{\textsf{\textbf{Text}}}
& \multicolumn{1}{c}{\textsf{\textbf{Image}}}
& \multicolumn{1}{c}{\textsf{\textbf{Video}}}
& \multicolumn{1}{c}{\textsf{\textbf{Audio}}}
\\[-2pt]

& \multicolumn{1}{c}{\odtasks{199}}
& \multicolumn{1}{c}{\odtasks{89}}
& \multicolumn{1}{c}{\odtasks{71}}
& \multicolumn{1}{c}{\odtasks{32}}
\\

\cmidrule(lr){2-2}
\cmidrule(lr){3-3}
\cmidrule(lr){4-4}
\cmidrule(lr){5-5}

\textbf{Model}
& Acc (\%)
& Acc (\%)
& Acc (\%)
& Acc (\%)
\\
\midrule

\modellogo{openai.png}\hspace{0.35em}GPT-6 Astra
& \odbest{38.69}{77}
& \odbest{48.31}{43}
& \odacc{40.85}{29}
& \odacc{43.75}{14}
\\

\modellogo{openai.png}\hspace{0.35em}GPT-5.6 Sol
& \odacc{28.14}{56}
& \odacc{32.58}{29}
& \odacc{29.58}{21}
& \odacc{25.00}{8}
\\

\modellogo{claude.png}\hspace{0.35em}Claude Opus 5.5
& \odacc{37.69}{75}
& \odacc{44.94}{40}
& \odacc{36.62}{26}
& \odacc{46.88}{15}
\\

\modellogo{claude.png}\hspace{0.35em}Claude Fable 5.1
& \odacc{33.67}{67}
& \odacc{38.20}{34}
& \odacc{32.39}{23}
& \odbest{50.00}{16}
\\

\modellogo{gemini.png}\hspace{0.35em}Gemini 3.8 Flash
& \odacc{36.68}{73}
& \odacc{31.46}{28}
& \odbest{43.66}{31}
& \odbest{50.00}{16}
\\

\modellogo{qwen.png}\hspace{0.35em}Qwen3.8-Max
& \odacc{25.63}{51}
& \odacc{25.84}{23}
& \odacc{30.99}{22}
& \odacc{34.38}{11}
\\

\modellogo{qwen.png}\hspace{0.35em}Qwen3.8-Flash
& \odacc{12.06}{24}
& \odacc{10.11}{9}
& \odacc{15.49}{11}
& \odacc{12.50}{4}
\\

\modellogo{deepseek.png}\hspace{0.35em}DeepSeek-V4.1-Flash
& \odacc{20.60}{41}
& \odacc{20.22}{18}
& \odacc{23.94}{17}
& \odacc{25.00}{8}
\\

\modellogo{deepseek.png}\hspace{0.35em}DeepSeek-V4-Pro
\textsuperscript{\textbf{(T)}}
& \odacc{12.56}{25}
& \odacc{8.99}{8}
& \odacc{12.68}{9}
& \odacc{15.62}{5}
\\

\modellogo{kimi.png}\hspace{0.35em}Kimi K3
& \odacc{17.59}{35}
& \odacc{16.85}{15}
& \odacc{25.35}{18}
& \odacc{31.25}{10}
\\

\modellogo{zai.png}\hspace{0.35em}GLM-5.3-Flash
& \odacc{23.62}{47}
& \odacc{25.84}{23}
& \odacc{28.17}{20}
& \odacc{34.38}{11}
\\

\modellogo{zai.png}\hspace{0.35em}GLM-5.3
\textsuperscript{\textbf{(T)}}
& \odacc{13.57}{27}
& \odacc{12.36}{11}
& \odacc{12.68}{9}
& \odacc{25.00}{8}
\\

\bottomrule
\end{tabular}%
}
\endgroup
\end{table}

\begin{table}[t]
\centering
\caption{\textbf{Results by data format on OpenDS-World.}
We report accuracy (\textbf{Acc}, \%) across eight resource format categories.
Numbers in parentheses indicate correct answers;
bold accuracy values denote the best result in each format.
A task may belong to multiple format categories.}
\label{tab:fit-format}

\begingroup
\small
\renewcommand{\arraystretch}{1.26}
\setlength{\tabcolsep}{4.2pt}
\setlength{\aboverulesep}{3pt}
\setlength{\belowrulesep}{3pt}
\setlength{\heavyrulewidth}{0.8pt}
\setlength{\lightrulewidth}{0.4pt}

\resizebox{0.88\textwidth}{!}{%
\begin{tabular}{
@{}l
r
@{\hspace{8pt}}r
@{\hspace{8pt}}r
@{\hspace{8pt}}r
@{\hspace{8pt}}r
@{\hspace{8pt}}r
@{\hspace{8pt}}r
@{\hspace{8pt}}r
@{}
}
\toprule

& \multicolumn{1}{c}{\textsf{\textbf{HTML}}}
& \multicolumn{1}{c}{\textsf{\textbf{MP4}}}
& \multicolumn{1}{c}{\textsf{\textbf{PDF}}}
& \multicolumn{1}{c}{\textsf{\textbf{MP3}}}
& \multicolumn{1}{c}{\textsf{\textbf{JPG}}}
& \multicolumn{1}{c}{\textsf{\textbf{CSV}}}
& \multicolumn{1}{c}{\textsf{\textbf{ZIP}}}
& \multicolumn{1}{c}{\textsf{\textbf{Other}}}
\\[-2pt]

& \multicolumn{1}{c}{\odtasks{196}}
& \multicolumn{1}{c}{\odtasks{70}}
& \multicolumn{1}{c}{\odtasks{62}}
& \multicolumn{1}{c}{\odtasks{31}}
& \multicolumn{1}{c}{\odtasks{19}}
& \multicolumn{1}{c}{\odtasks{15}}
& \multicolumn{1}{c}{\odtasks{14}}
& \multicolumn{1}{c}{\odtasks{79}}
\\

\cmidrule(lr){2-2}
\cmidrule(lr){3-3}
\cmidrule(lr){4-4}
\cmidrule(lr){5-5}
\cmidrule(lr){6-6}
\cmidrule(lr){7-7}
\cmidrule(lr){8-8}
\cmidrule(lr){9-9}

\textbf{Model}
& Acc (\%)
& Acc (\%)
& Acc (\%)
& Acc (\%)
& Acc (\%)
& Acc (\%)
& Acc (\%)
& Acc (\%)
\\
\midrule

\modellogo{openai.png}\hspace{0.35em}GPT-6 Astra
& \odbest{39.29}{77}
& \odacc{40.00}{28}
& \odbest{50.00}{31}
& \odacc{45.16}{14}
& \odbest{31.58}{6}
& \odbest{60.00}{9}
& \odbest{57.14}{8}
& \odbest{41.77}{33}
\\

\modellogo{openai.png}\hspace{0.35em}GPT-5.6 Sol
& \odacc{28.06}{55}
& \odacc{28.57}{20}
& \odacc{35.48}{22}
& \odacc{25.81}{8}
& \odacc{21.05}{4}
& \odacc{26.67}{4}
& \odacc{7.14}{1}
& \odacc{25.32}{20}
\\

\modellogo{claude.png}\hspace{0.35em}Claude Opus 5.5
& \odacc{38.27}{75}
& \odacc{35.71}{25}
& \odacc{46.77}{29}
& \odacc{45.16}{14}
& \odacc{26.32}{5}
& \odacc{53.33}{8}
& \odacc{35.71}{5}
& \odacc{39.24}{31}
\\

\modellogo{claude.png}\hspace{0.35em}Claude Fable 5.1
& \odacc{33.16}{65}
& \odacc{31.43}{22}
& \odacc{37.10}{23}
& \odacc{48.39}{15}
& \odacc{15.79}{3}
& \odacc{46.67}{7}
& \odacc{21.43}{3}
& \odacc{40.51}{32}
\\

\modellogo{gemini.png}\hspace{0.35em}Gemini 3.8 Flash
& \odacc{36.73}{72}
& \odbest{42.86}{30}
& \odacc{37.10}{23}
& \odbest{51.61}{16}
& \odacc{21.05}{4}
& \odacc{33.33}{5}
& \odacc{28.57}{4}
& \odacc{31.65}{25}
\\

\modellogo{qwen.png}\hspace{0.35em}Qwen3.8-Max
& \odacc{26.02}{51}
& \odacc{30.00}{21}
& \odacc{24.19}{15}
& \odacc{35.48}{11}
& \odacc{21.05}{4}
& \odacc{53.33}{8}
& \odacc{14.29}{2}
& \odacc{26.58}{21}
\\

\modellogo{qwen.png}\hspace{0.35em}Qwen3.8-Flash
& \odacc{12.24}{24}
& \odacc{14.29}{10}
& \odacc{8.06}{5}
& \odacc{12.90}{4}
& \odacc{10.53}{2}
& \odacc{33.33}{5}
& \odacc{0.00}{0}
& \odacc{10.13}{8}
\\

\modellogo{deepseek.png}\hspace{0.35em}DeepSeek-V4.1-Flash
& \odacc{20.92}{41}
& \odacc{22.86}{16}
& \odacc{17.74}{11}
& \odacc{25.81}{8}
& \odacc{10.53}{2}
& \odacc{33.33}{5}
& \odacc{14.29}{2}
& \odacc{18.99}{15}
\\

\modellogo{deepseek.png}\hspace{0.35em}DeepSeek-V4-Pro
\textsuperscript{\textbf{(T)}}
& \odacc{12.24}{24}
& \odacc{11.43}{8}
& \odacc{12.90}{8}
& \odacc{16.13}{5}
& \odacc{5.26}{1}
& \odacc{26.67}{4}
& \odacc{0.00}{0}
& \odacc{12.66}{10}
\\

\modellogo{kimi.png}\hspace{0.35em}Kimi K3
& \odacc{17.35}{34}
& \odacc{24.29}{17}
& \odacc{11.29}{7}
& \odacc{32.26}{10}
& \odacc{0.00}{0}
& \odacc{26.67}{4}
& \odacc{14.29}{2}
& \odacc{20.25}{16}
\\

\modellogo{zai.png}\hspace{0.35em}GLM-5.3-Flash
& \odacc{23.98}{47}
& \odacc{27.14}{19}
& \odacc{19.35}{12}
& \odacc{35.48}{11}
& \odacc{5.26}{1}
& \odacc{26.67}{4}
& \odacc{21.43}{3}
& \odacc{26.58}{21}
\\

\modellogo{zai.png}\hspace{0.35em}GLM-5.3
\textsuperscript{\textbf{(T)}}
& \odacc{13.27}{26}
& \odacc{11.43}{8}
& \odacc{16.13}{10}
& \odacc{25.81}{8}
& \odacc{0.00}{0}
& \odacc{26.67}{4}
& \odacc{14.29}{2}
& \odacc{11.39}{9}
\\

\bottomrule
\end{tabular}%
}
\endgroup
\end{table}

\begin{table}[t]
\centering
\caption{\textbf{Results by access requirement on OpenDS-World.}
We report accuracy (\textbf{Acc}, \%) across seven access-requirement categories.
Numbers in parentheses indicate correct answers;
bold accuracy values denote the best result in each category.}
\label{tab:fit-access}

\begingroup
\small
\renewcommand{\arraystretch}{1.26}
\setlength{\tabcolsep}{4.2pt}
\setlength{\aboverulesep}{3pt}
\setlength{\belowrulesep}{3pt}
\setlength{\heavyrulewidth}{0.8pt}
\setlength{\lightrulewidth}{0.4pt}

\resizebox{0.88\textwidth}{!}{%
\begin{tabular}{
@{}l
r
@{\hspace{10pt}}r
@{\hspace{10pt}}r
@{\hspace{10pt}}r
@{\hspace{10pt}}r
@{\hspace{10pt}}r
@{\hspace{10pt}}r
@{}
}
\toprule

& \multicolumn{1}{c}{\textsf{\textbf{O}}}
& \multicolumn{1}{c}{\textsf{\textbf{N}}}
& \multicolumn{1}{c}{\textsf{\textbf{X}}}
& \multicolumn{1}{c}{\textsf{\textbf{T}}}
& \multicolumn{1}{c}{\textsf{\textbf{P}}}
& \multicolumn{1}{c}{\textsf{\textbf{G}}}
& \multicolumn{1}{c}{\textsf{\textbf{Q}}}
\\[-2pt]

& \multicolumn{1}{c}{\odtasks{200}}
& \multicolumn{1}{c}{\odtasks{134}}
& \multicolumn{1}{c}{\odtasks{120}}
& \multicolumn{1}{c}{\odtasks{95}}
& \multicolumn{1}{c}{\odtasks{84}}
& \multicolumn{1}{c}{\odtasks{25}}
& \multicolumn{1}{c}{\odtasks{13}}
\\

\cmidrule(lr){2-2}
\cmidrule(lr){3-3}
\cmidrule(lr){4-4}
\cmidrule(lr){5-5}
\cmidrule(lr){6-6}
\cmidrule(lr){7-7}
\cmidrule(lr){8-8}

\textbf{Model}
& Acc (\%)
& Acc (\%)
& Acc (\%)
& Acc (\%)
& Acc (\%)
& Acc (\%)
& Acc (\%)
\\
\midrule

\modellogo{openai.png}\hspace{0.35em}GPT-6 Astra
& \odbest{39.00}{78}
& \odbest{41.79}{56}
& \odbest{46.67}{56}
& \odacc{40.00}{38}
& \odbest{42.86}{36}
& \odbest{40.00}{10}
& \odbest{61.54}{8}
\\

\modellogo{openai.png}\hspace{0.35em}GPT-5.6 Sol
& \odacc{28.00}{56}
& \odacc{27.61}{37}
& \odacc{29.17}{35}
& \odacc{28.42}{27}
& \odacc{32.14}{27}
& \odacc{24.00}{6}
& \odacc{38.46}{5}
\\

\modellogo{claude.png}\hspace{0.35em}Claude Opus 5.5
& \odacc{38.00}{76}
& \odacc{38.06}{51}
& \odacc{43.33}{52}
& \odacc{40.00}{38}
& \odacc{41.67}{35}
& \odacc{28.00}{7}
& \odacc{53.85}{7}
\\

\modellogo{claude.png}\hspace{0.35em}Claude Fable 5.1
& \odacc{33.50}{67}
& \odacc{34.33}{46}
& \odacc{40.00}{48}
& \odacc{38.95}{37}
& \odacc{38.10}{32}
& \odacc{32.00}{8}
& \odacc{46.15}{6}
\\

\modellogo{gemini.png}\hspace{0.35em}Gemini 3.8 Flash
& \odacc{36.50}{73}
& \odacc{33.58}{45}
& \odacc{37.50}{45}
& \odbest{46.32}{44}
& \odacc{34.52}{29}
& \odacc{28.00}{7}
& \odacc{53.85}{7}
\\

\modellogo{qwen.png}\hspace{0.35em}Qwen3.8-Max
& \odacc{25.50}{51}
& \odacc{23.88}{32}
& \odacc{28.33}{34}
& \odacc{32.63}{31}
& \odacc{23.81}{20}
& \odacc{24.00}{6}
& \odacc{38.46}{5}
\\

\modellogo{qwen.png}\hspace{0.35em}Qwen3.8-Flash
& \odacc{12.00}{24}
& \odacc{8.21}{11}
& \odacc{10.83}{13}
& \odacc{14.74}{14}
& \odacc{8.33}{7}
& \odacc{8.00}{2}
& \odacc{23.08}{3}
\\

\modellogo{deepseek.png}\hspace{0.35em}DeepSeek-V4.1-Flash
& \odacc{20.50}{41}
& \odacc{17.91}{24}
& \odacc{20.83}{25}
& \odacc{24.21}{23}
& \odacc{17.86}{15}
& \odacc{12.00}{3}
& \odacc{30.77}{4}
\\

\modellogo{deepseek.png}\hspace{0.35em}DeepSeek-V4-Pro
\textsuperscript{\textbf{(T)}}
& \odacc{12.50}{25}
& \odacc{10.45}{14}
& \odacc{14.17}{17}
& \odacc{13.68}{13}
& \odacc{7.14}{6}
& \odacc{4.00}{1}
& \odacc{23.08}{3}
\\

\modellogo{kimi.png}\hspace{0.35em}Kimi K3
& \odacc{17.50}{35}
& \odacc{15.67}{21}
& \odacc{19.17}{23}
& \odacc{28.42}{27}
& \odacc{19.05}{16}
& \odacc{20.00}{5}
& \odacc{0.00}{0}
\\

\modellogo{zai.png}\hspace{0.35em}GLM-5.3-Flash
& \odacc{23.50}{47}
& \odacc{22.39}{30}
& \odacc{24.17}{29}
& \odacc{29.47}{28}
& \odacc{21.43}{18}
& \odacc{24.00}{6}
& \odacc{15.38}{2}
\\

\modellogo{zai.png}\hspace{0.35em}GLM-5.3
\textsuperscript{\textbf{(T)}}
& \odacc{13.50}{27}
& \odacc{12.69}{17}
& \odacc{15.00}{18}
& \odacc{15.79}{15}
& \odacc{14.29}{12}
& \odacc{12.00}{3}
& \odacc{15.38}{2}
\\

\bottomrule
\end{tabular}%
}
\endgroup
\end{table}

\paragraph{Data Format.}
Table~\ref{tab:fit-format} presents the accuracy of 12 models across eight data format categories: HTML (196 tasks), MP4 (70), PDF (62), MP3 (31), JPG (19), CSV (15), ZIP (14), and Other (79). GPT-6 Astra achieves the highest accuracy in six categories, including PDF (50.00\%), CSV (60.00\%), and ZIP (57.14\%). Gemini 3.8 Flash leads in MP4 (42.86\%) and MP3 (51.61\%), highlighting its relative advantage in audiovisual formats. Claude Opus 5.5 performs strongly on PDF (46.77\%) and CSV (53.33\%), while Claude Fable 5.1 achieves 48.39\% on MP3 and 40.51\% on Other formats. GPT-5.6 Sol also records its highest format-specific accuracy on PDF (35.48\%). Among the remaining models, CSV exhibits notably strong performance across several model families. Qwen3.8-Max achieves 53.33\% accuracy on CSV, while Qwen3.8-Flash and DeepSeek-V4.1-Flash both reach 33.33\%. DeepSeek-V4-Pro (T) and GLM-5.3 (T) also achieve their highest format-specific accuracies on CSV (both 26.67\%). In contrast, Kimi K3 and GLM-5.3-Flash perform relatively well on MP3, reaching 32.26\% and 35.48\%, respectively. Across models, JPG generally exhibits lower accuracy, while ZIP shows substantial performance variation, ranging from 0.00\% to 57.14\%. Overall, the results reveal distinct performance patterns across data formats, with GPT-6 Astra leading in most categories, Gemini 3.8 Flash excelling in audiovisual formats, and several other models exhibiting relative strengths in structured-data processing.

\paragraph{Access Requirements.} Table~\ref{tab:fit-access} presents the accuracy of 12 models across seven access requirement categories. GPT-6 Astra achieves the highest accuracy in six categories, including Artifact Transfer (X) and Declarative Query (Q), with accuracies of 46.67\% and 61.54\%, respectively. Gemini 3.8 Flash leads in Temporal Indexing (T) with 46.32\% accuracy, outperforming GPT-6 Astra and Claude Opus 5.5 (both 40.00\%). Claude Opus 5.5 and Claude Fable 5.1 achieve 43.33\% and 40.00\% accuracy in Artifact Transfer, respectively, while GPT-5.6 Sol reaches 32.14\% in Parameterized State Update (P). Among the remaining models, Temporal Indexing exhibits a relatively consistent performance pattern. Qwen3.8-Max, GLM-5.3-Flash, and Kimi K3 achieve 32.63\%, 29.47\%, and 28.42\% accuracy, respectively, all exceeding their corresponding overall accuracies. DeepSeek-V4.1-Flash also achieves relatively high accuracy in Temporal Indexing (24.21\%) and Declarative Query (30.77\%). In contrast, Qwen3.8-Flash, DeepSeek-V4-Pro (T), and GLM-5.3 (T) exhibit relatively low accuracy across most access requirement categories. Across the 12 models, Declarative Query (Q) exhibits pronounced variation in accuracy, ranging from 0.00\% to 61.54\%. These results highlight performance differences among models across access-requirement categories.

\subsection{Video Duration Analysis}
\label{sec:more-length}
\begin{figure}[tbp]
\centering
\includegraphics[width=0.85\linewidth]{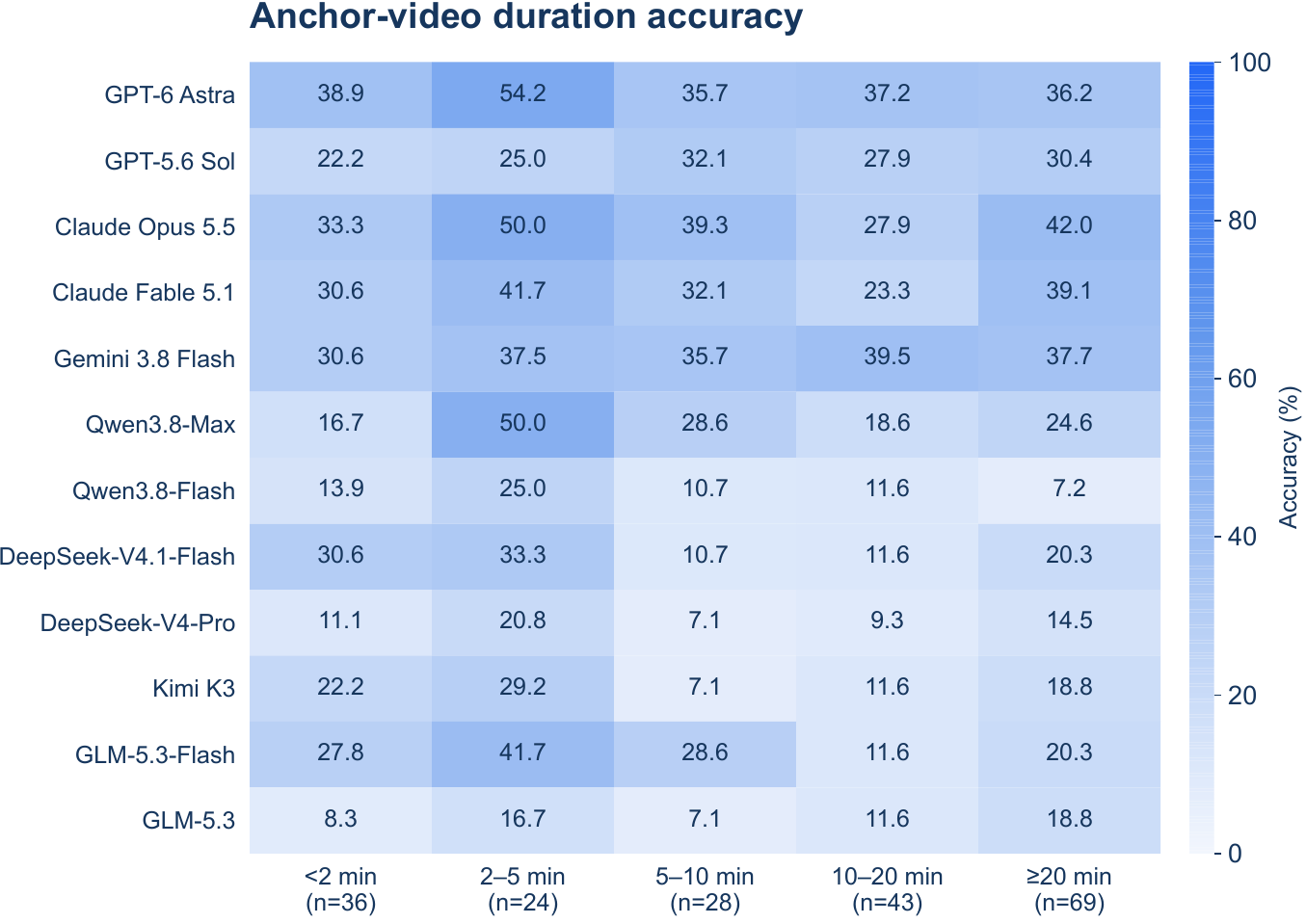}
\caption{Model accuracy across video-duration groups on OpenDS-World. Rows represent the 12 evaluated models, and columns correspond to five video-duration intervals. Values indicate accuracy (\%), with the number of tasks in each interval shown in parentheses.}
\label{fig:fit-length}
\end{figure}

Figure~\ref{fig:fit-length} presents model accuracy across five anchor-video-duration groups: [0, 2), [2, 5), [5, 10), [10, 20), and [20, $\infty$) minutes, containing 36, 24, 28, 43, and 69 tasks, respectively. These groups are shared by all 12 evaluated models. The analysis considers the duration of the provided anchor video. Performance varies across duration groups, but the observed differences do not establish a causal relationship between video duration and task difficulty.

\section{Discussion}
\label{sec:discussion}

\begin{figure}[tbp]
\centering
\includegraphics[width=0.85\linewidth]{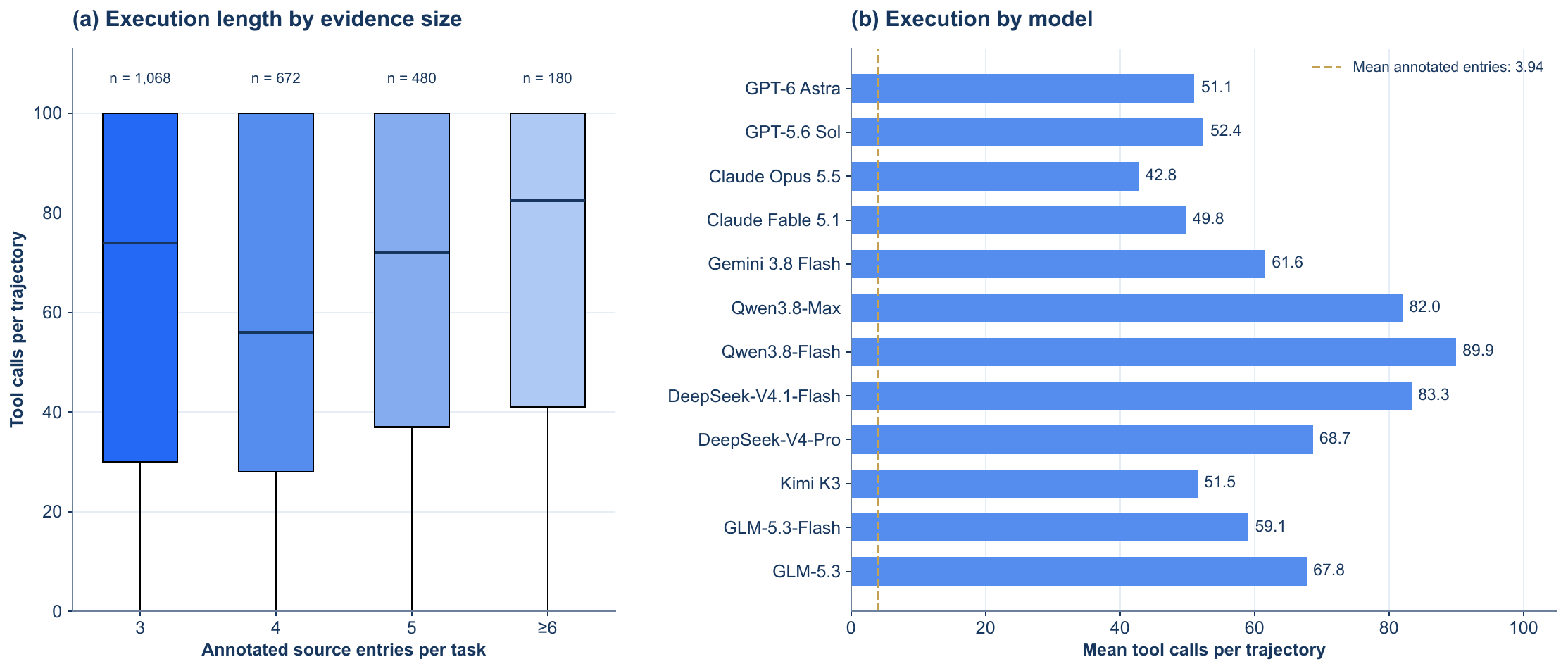}
\caption{\textbf{Long-horizon execution in OpenDS-World.} (a) Distribution of tool-call counts across observed trajectories grouped by the number of annotated source entries. Boxes indicate interquartile ranges with median lines; outliers are omitted. (b) Mean tool calls per observed trajectory for each model, compared with the average number of annotated source entries per task (3.94).}
\label{fig:long-horizon}
\end{figure}

\paragraph{Natural Long-Horizon Tasks.}
\label{sec:long-horizon-tasks}

Deep search tasks in OpenDS-World inherently exhibit long-horizon decision-making characteristics. Unlike conventional single-round retrieval or short-horizon question answering, these tasks require agents to identify key clues from the initial video and progressively discover, access, and integrate evidence distributed across multiple sources in the open Web. This process involves iterative querying, cross-page navigation, interactive information access, and multimodal evidence verification, resulting in execution trajectories substantially longer than the annotated evidence chains. Figure~\ref{fig:long-horizon} provides empirical evidence of this distinction. As shown in Figure~\ref{fig:long-horizon}(a), tasks with 3, 4, 5, and at least 6 annotated source entries exhibit median tool-call counts of 74, 56, 72, and 82.5, respectively. Even tasks requiring only three annotated sources involve extensive tool interactions, highlighting the substantial exploration required to locate and verify the final evidence. Moreover, execution length does not increase monotonically with evidence-chain size, indicating that the number of annotated sources alone does not determine the complexity of the research process. Figure~\ref{fig:long-horizon}(b) further shows that the mean number of tool calls varies from 42.8 to 89.9 across models, substantially exceeding the average of 3.94 annotated source entries per task. The concentration of trajectories near the 100-call budget also reveals the influence of execution constraints on observed search behavior. Together, these findings demonstrate that compact evidence chains can require extensive exploration and interaction, making long-horizon planning, context management, and adaptive tool use central challenges in real-world deep search~\citep{gao2025beyond, wu2025resum, qiao2025webresearcher}.

\begin{figure}[tbp]
\centering
\includegraphics[width=0.98\linewidth]{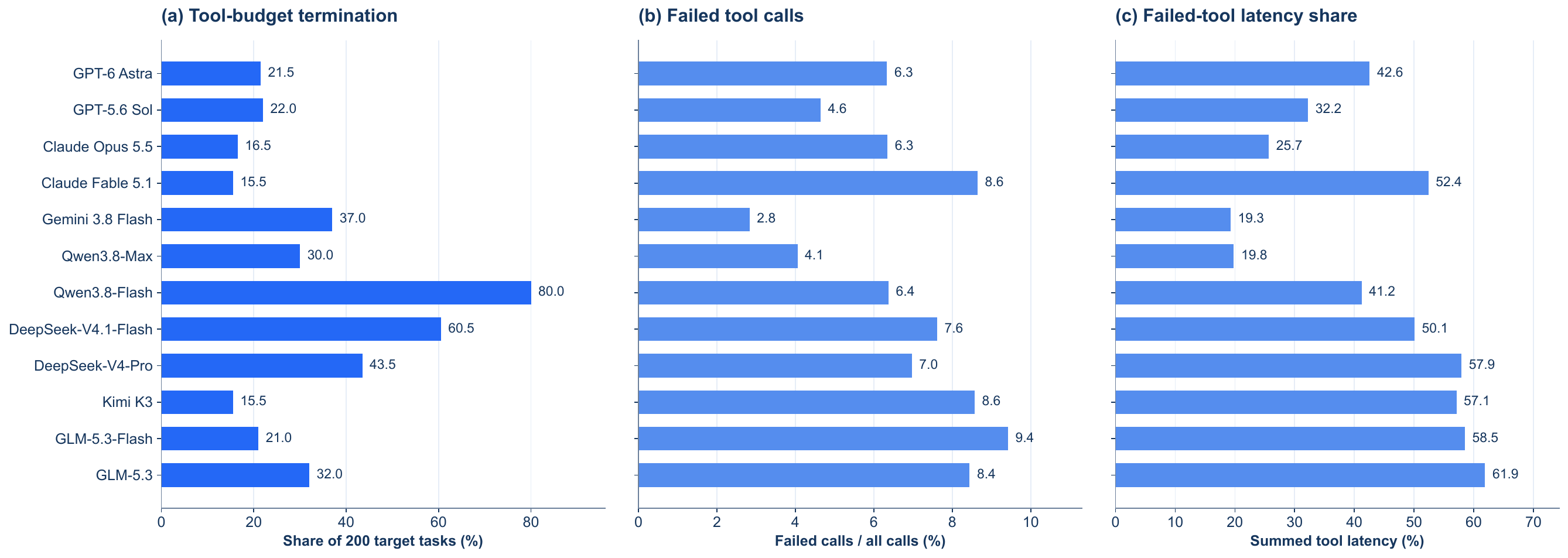}
\caption{\textbf{Execution-level opportunities for agent improvement.}
(a) Percentage of the 200 target tasks terminated by tool-call budget exhaustion.
(b) Percentage of failed tool calls among all recorded tool calls.
(c) Percentage of total recorded tool-execution latency attributed to failed calls.
These statistics characterize execution inefficiencies and potential targets for adaptive strategy improvement.}
\label{fig:improvement-opportunities}
\end{figure}

\paragraph{A Natural Testbed for Recursive Self-Improvement}
\label{sec:recursive-self-improvement}

OpenDS-World provides a realistic, open, and dynamic Web environment for studying Recursive Self-Improvement (RSI). Unlike benchmarks based on static corpora or predefined retrieval paths, OpenDS-World requires agents to autonomously navigate changing search results, heterogeneous Web interfaces,  and interactive resources while handling uncertain tool outcomes. These characteristics expose agents to diverse execution failures and generate actionable environmental feedback, creating opportunities to improve search strategies, tool selection, error recovery, and resource allocation through accumulated experience~\citep{zheng2025deepresearcher}. Figure~\ref{fig:improvement-opportunities} quantifies several of these improvement opportunities. As shown in Figure~\ref{fig:improvement-opportunities}(a), the proportion of trajectories terminated by tool-budget exhaustion ranges from 15.5\% to 80.0\% across models, with Qwen3.8-Flash reaching 80.0\% and DeepSeek-V4.1-Flash reaching 60.5\%. This substantial variation highlights the importance of adaptive exploration and termination strategies~\citep{guan2026deeprag} under constrained interaction budgets. Figures~\ref{fig:improvement-opportunities}(b) and (c) further characterize unsuccessful tool operations and their associated execution costs. Notably, failed tool calls account for 19.3\% to 61.9\% of cumulative tool-execution latency across models, indicating that unsuccessful interactions can consume a substantial portion of the available execution time. These observations identify concrete optimization targets, including reducing repeated failures, improving recovery decisions, and prioritizing more effective interactions. Beyond these execution-level signals, OpenDS-World provides 200 tasks with verifiable answers, 787 verified source entries, and heterogeneous evidence spanning 42 file formats and seven access requirements. The combination of answer-level verification, detailed execution trajectories, and recurring interaction challenges establishes a structured feedback environment for self-improvement. Agents can identify recurring failure patterns from historical trajectories, derive transferable strategies, and evaluate  revisions on new tasks. Importantly, the dynamic nature of the real Web requires such improvements to generalize beyond fixed task instances and predefined interaction sequences. Therefore, OpenDS-World offers a natural testbed for investigating how autonomous agents can progressively refine their research and decision-making mechanisms through verifiable feedback from real-world environments.





\end{document}